%% file: iclr2026_conference.tex
\documentclass{article}

\newif\ifarxiv
\arxivtrue      

\ifarxiv
  \usepackage[margin=1in]{geometry}
  \usepackage[numbers,sort&compress]{natbib}
\else
  \usepackage{iclr2026_conference,times}
\fi

\input{math_commands.tex}

\usepackage[section]{placeins} 
\usepackage{subcaption}

\usepackage{tikz}
\usetikzlibrary{arrows.meta,positioning,calc,decorations.pathreplacing}

\DeclareMathOperator{\Exp}{Exp}

\usepackage{amsthm,amsmath,amssymb}
\usepackage{url}
\usepackage{algorithm}
\usepackage[noend]{algpseudocode}
\usepackage{booktabs}
\usepackage{xcolor}
\usepackage{microtype}
\usepackage{graphicx}     

\newtheorem{theorem}{Theorem}
\newtheorem{lemma}{Lemma}
\newtheorem{assumption}{Assumption}
\newtheorem{proposition}{Proposition}

\newtheorem{definition}{Definition}

\newcommand{\Key}{\mathrm{Key}}
\newcommand{\Keyhat}{\widehat{\mathrm{Key}}}
\newcommand{\Emin}{E_{\min}}
\newcommand{\Nub}{N_{\mathrm{ub}}}
\usepackage{hyperref}
\ifarxiv
  \hypersetup{hidelinks,
              pdftitle={Auditable Early Stopping for Agentic Routing: Ledger-Verified Run-Wise Certificates under Local DP},
              pdfauthor={Shivam Akhauri}}
\else
  \hypersetup{hidelinks}
\fi

\newenvironment{boxedremark}
  {\begin{center}\begin{minipage}{0.97\linewidth}\hrule\smallskip}
  {\smallskip\hrule\end{minipage}\end{center}}

\title{Auditable Early Stopping for Agentic Routing: Ledger-Verified Run-Wise Certificates under Local DP}

\ifarxiv
\author{Shivam Akhauri \\
\texttt{shivamakhauri04@gmail.com}}
\else
\author{Anonymous Authors\\
\texttt{[[anonymized for review]]}}
\fi

\begin{document}
\maketitle

\begin{abstract}
We study early stopping for best-first routing in tool-use agents under local differential privacy (LDP), with an auditable, validator-replayable ledger. Our key idea is a \emph{run-wise certificate}: we couple each node's key to the \emph{same} exponential race that realizes leaf perturbations, so the standard halting rule---stop when $\max_{v\in\mathcal{F}}\Key(v)\le B^{\ast}$, where $B^{\ast}$ is the incumbent realized leaf value---soundly certifies the realized run. We provide two certified modes on context-indexed prefix--DAGs whose children partition the leaf set. \emph{Exact} mode (known counts) implements lazy offset propagation with winner reuse; \emph{Surrogate} mode (upper bounds only) disables winner reuse and uses a parent-anchored surrogate race; keys are conservative online and can be tightened ex post by a validator via $\kappa=\log(N/\Nub)$. A small compiler enforces the partition property, and an admissible, race-independent $M_\tau$ keeps keys sound. A replayable ledger records uniforms, counts, and tie handling; privacy follows from post-processing. Experiments on synthetic graphs and a small real tool-use pipeline show tight stopping, deterministic replay, and low overhead.
\end{abstract}

\section{One-Minute Story (Running Example)}
You route a question through a tool pipeline (retrieval $\to$ summarization $\to$ calculator) under \textbf{local DP}: inputs are privatized upstream; routing must be auditable yet privacy-preserving. Your grammar yields a \emph{shared-node} DAG (subtools reused across contexts). Our compiler \emph{context-indexes} nodes to form a \emph{prefix--DAG} whose children \emph{partition} descendant leaves. The router runs \emph{Lazy--A* PaM} with realized path score
\[
s_{\mathrm{det}}(P)=-\,C(P)\qquad\text{(certified deterministic score; PaM/Gumbel may be used as a heuristic only)}
\]
we log the leaf \emph{uniforms} $U(P)$ (and the node-level uniforms used to lazy-simulate the race) so the keys are reconstructible. If PaM/Gumbel is used for \emph{heuristic} ranking (marked \texttt{NoCert}), we also reconstruct $G(P)$ from the same $U(P)$; heuristics never affect the certified keys. The priority-queue (PQ) key at node $v$ is $\Key(v)=M_\tau(v)-\log t(v)$, where $t(v)=\min_{P\in\mathcal P(v)}E_P$ for i.i.d.\ $E_P\!\sim\!\mathrm{Exp}(1)$ (single \emph{exponential race} with lazy offset propagation). With exact counts $N$ you get \emph{Exact mode} lazy reuse; with only \emph{upper bounds} $\Nub\!\ge\!N$, you switch to \emph{Surrogate mode}: \emph{no winner reuse}, and all child keys use a parent-anchored surrogate race $\hat t(\cdot)$, keeping keys safe since $-\log\hat t\!\ge\!-\log t$. The ledger records $U(P)$, $\Nub$, and claim types; where public counts are available, the validator computes $\kappa=\log(N/\Nub)$ to \emph{tighten} keys, otherwise keys remain conservative. A potential $\Phi$ guarantees acyclicity; truncation is certified in \emph{score units}. Every prune is \emph{replayable}. Modern tool-use agents motivate such routing under constraints \citep{Schick2023Toolformer,Yao2023ReAct,Patil2024Gorilla}.

\paragraph{Contributions.}
\begin{itemize}
\item \textbf{Run-wise coupling.} We couple pruning keys to the \emph{same} exponential race that realizes leaf perturbations, yielding a frontier-sound halting rule: stop when $\max_{v\in\mathcal{F}}\Key(v)\le B^{\ast}$ on context-indexed prefix--DAGs.
\item \textbf{Two certified modes.} \emph{Exact} (winner reuse via lazy offset propagation) and \emph{Surrogate} (no winner reuse; parent-anchored keys), with validator-side tightening via $\kappa=\log(N/\Nub)$.
\item \textbf{Deployment artifacts.} An audit-ready ledger/validator protocol and a small compiler enforcing child partition. Experiments on synthetic graphs and a small tool-use pipeline show tight stopping, deterministic replay, and low overhead.
\end{itemize}

\section{Model, Assumptions, and Notation}

\paragraph{Prefix--DAG \& child partition.}
The search structure is a \emph{context-indexed prefix--DAG} $\mathcal G=(V,E)$ (formal defs in Appx.~\ref{app:compiler-transform}): each node $v$ encodes a tuple (tool, state, grammar version, \emph{prefix context}), i.e., the root-to-$v$ prefix uniquely determines which leaves it can reach. Let $\mathcal P(v)$ denote the set of \emph{canonical leaf identifiers} reachable from $v$ after context indexing, and write $N(v):=|\mathcal P(v)|$.

\textbf{Child-partition property.} For any internal $v$ with children $\mathrm{Ch}(v)$, the sets $\{\mathcal P(u)\}_{u\in\mathrm{Ch}(v)}$ are pairwise disjoint and
\[
\biguplus_{u\in \mathrm{Ch}(v)} \mathcal P(u) \;=\; \mathcal P(v),
\]
so in particular $N(v)=\sum_{u\in\mathrm{Ch}(v)} N(u)$.

\paragraph{Local finiteness.}
\begin{assumption}[Local finiteness]\label{as:local-finiteness}
For every explored node $v$ (i.e., any node generated/inserted into the PQ), the number of descendant leaves is finite: $N(v)<\infty$.
\end{assumption}
Under Assumption~\ref{as:local-finiteness}, the minimum arrival time
\[
t(v)\;:=\;\min_{P\in\mathcal P(v)} E_P,\qquad E_P\stackrel{\text{i.i.d.}}{\sim}\mathrm{Exp}(1),
\]
is strictly positive almost surely, so $-\log t(v)$ is well defined.
The compiler emits a machine-checkable \emph{finiteness/partition certificate}; if certification fails or counters overflow/timeout at runtime, the router \emph{mandatorily downgrades to Surrogate mode} (disables winner reuse and uses parent-anchored surrogate races) to preserve soundness.

\paragraph{Race, minima, keys, uniforms.}
Each leaf $P$ is assigned a uniform $U_P\in(0,1)$. In \textbf{Exact} mode we do not resample at leaf pop—$U_P$ is recovered deterministically via $U_P=1-e^{-t(P)}$ from the realized race. In \textbf{Surrogate} mode the true per‑leaf arrival is not realized by the surrogate race, so at leaf pop we \emph{draw or PRF‑derive} a fresh $U_P$ and set $E_P=-\log(1-U_P)$. (In \textbf{Fallback} $U_P$ also comes from a PRF.) We keep the standard couplings for replay:
\[
E_P=-\log\!\bigl(1-U_P\bigr)\ \sim\ \mathrm{Exp}(1),
\qquad 
G(P)=-\log\!\bigl(-\log U_P\bigr)\ \sim\ \mathrm{Gumbel}(0,1).
\]

\footnote{We map unsigned 64-bit integers $x\in\{0,\ldots,2^{64}-1\}$ to an open-interval uniform $U=(x+0.5)\,2^{-64}\in(0,1)$, so $\log$ and $\log\!\log$ never see endpoints; the $+0.5$ is midpoint quantization. See \citep{Reynolds2020UniformFloat}. In code, compute $E_P$ via \texttt{-log1p(-U)} for numerical stability.}
For node $v$, set
\[
t(v):=\min_{P\in\mathcal P(v)}E_P,\qquad 
\Key(v)=M_\tau(v)-\log t(v),\qquad 
\Keyhat(v)=M_\tau(v)-\log \hat t(v)\ \text{(Surrogate)}.
\]
For the \textbf{certified} objective we use $s_{\mathrm{det}}(P)=-C(P)$. If we additionally use PaM/Gumbel for \emph{heuristic} ranking (marked \texttt{NoCert}), we compute $G(P)$ from the same $U_P$; this does not affect keys or certificates.\footnote{A distribution-level baseline is $\mathbb{E}[-\log \Emin]=\gamma+\log N$ for $\Emin\!\sim\!\mathrm{Exp}(N)$, where $\gamma$ is Euler’s constant.}

\paragraph{Exact-mode coupling at leaves (no new randomness).}
In \textbf{Exact} mode, when a leaf $P$ is popped we do \emph{not} draw a fresh $U_P$.
Instead we take the leaf's arrival time from the already realized race: $E \leftarrow t(P)$. We then set
$U_P := 1 - e^{-E} \in (0,1)$ and compute $G(P) := -\log(-\log U_P)$. Thus both $E_P$ and $G(P)$ are measurable
with respect to the \emph{same} realized race. The ledger logs $U_P$ for replay, but $U_P$ is a deterministic
function of the race values already recorded.
\paragraph{Surrogate-mode leaf instantiation (fresh uniforms).}
In \textbf{Surrogate} mode the true per-leaf arrival times $E_P$ are not realized by the surrogate node race.
When a leaf $P$ is popped we draw (or PRF-derive) a fresh $U_P\in(0,1)$ and set $E_P:=-\log(1-U_P)$.
We then evaluate the certified augmented value $V(P)=s_{\mathrm{det}}(P)-\log E_P$ and log $U_P$ for replay.

\paragraph{Admissible, race-independent $M_\tau$.}
$M_\tau$ is deterministic (independent of the race uniforms) and upper-bounds the remaining score from $v$; it is computed by recipes R1/R2 (Sec.~\ref{sec:mtau}).

\paragraph{Stop rule, incumbent, and ties.}
Let $\mathcal F$ be the PQ frontier and define the \emph{augmented} leaf value
$V(P):=s_{\mathrm{det}}(P)-\log E_P$. We maintain the incumbent
$B^{*}=\max_{\text{expanded } P} V(P).$
We halt when either \textbf{(Exact)} $\max_{v\in\mathcal{F}} \Key(v) \le B^{*}$ or
\textbf{(Surrogate)} $\max_{v\in\mathcal{F}} \Keyhat(v) \le B^{*}$.
By Lemma~\ref{lem:frontier-surrogate}, these keys upper-bound realized subtree
maxima almost surely, so the corresponding stop rules are sound for the realized run.
With real-valued $U_P$ the race is a.s.\ tie-free; with fixed-point $U_P$ we log a
\texttt{tie\_token} and resolve ties lexicographically on public IDs, so the stop rule
remains a measurable function of the ledger (replayable by the validator), cf.\ \citep{Danihelka2022GumbelPlanning}.

\textbf{Realized subtree maximum:} $\mathrm{RSM}(v)=\max_{P\in\mathcal P(v)}\big(s_{\mathrm{det}}(P)-\log E_P\big)$.
\textbf{Run-wise certificate:} If at stop either $\max_{u\in\mathcal{F}} \Key(u) \le B^{*}$ (Exact) or
$\max_{u\in\mathcal{F}} \Keyhat(u) \le B^{*}$ (Surrogate), then every unexpanded leaf $P$
satisfies $s_{\mathrm{det}}(P)-\log E_P \le B^{*}$ under the \emph{same} realized race.

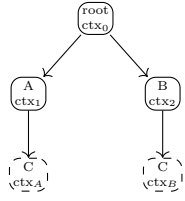
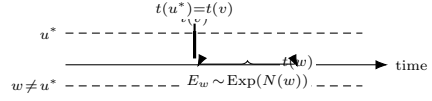
\begin{figure}[!t]
\centering
\begingroup

\begin{subfigure}[t]{0.46\linewidth}
\centering
\begin{tikzpicture}[
    scale=0.7, transform shape,
    node distance=6.5mm,
    every node/.style={font=\scriptsize},
    box/.style={draw, rounded corners, inner sep=2pt, align=center}
]
  \node[box] (r){root\\ctx$_0$};
  \node[box, below left=8mm and 6mm of r] (a){A\\ctx$_1$};
  \node[box, below right=8mm and 6mm of r] (b){B\\ctx$_2$};
  \node[box, dashed, below=9mm of a] (ca){C\\ctx$_A$};
  \node[box, dashed, below=9mm of b] (cb){C\\ctx$_B$};
  \draw[->] (r)--(a);
  \draw[->] (r)--(b);
  \draw[->] (a)--(ca);
  \draw[->] (b)--(cb);
\end{tikzpicture}
\caption{\textbf{Context indexing.} Diamonds are resolved by context: children partition the leaf IDs at every node (prefix--DAG child-partition).}
\label{fig:dag-left}
\end{subfigure}\hfill
\begin{subfigure}[t]{0.46\linewidth}
\centering
\begin{tikzpicture}[
    scale=0.78, transform shape,
    x=2.2cm, y=0.9cm,
    >=Latex,
    every node/.style={font=\scriptsize}
]
  \draw[->] (0,0) -- (2.5,0) node[right]{time};

  \draw[densely dashed] (0,0.60) -- (2.3,0.60);
  \draw[densely dashed] (0,-0.40) -- (2.3,-0.40);
  \node[anchor=east] at (0,0.60) {$u^{*}$};
  \node[anchor=east] at (0,-0.40) {$w\!\ne\!u^{*}$};

  \draw[very thick] (1.00,0.12) -- ++(0,0.52);
  \node[above, fill=white, inner sep=1pt] at (1.00,0.70) {$t(v)$};

  \draw[thick] (1.00,0.60) -- ++(0,0.18);
  \node[above, fill=white, inner sep=1pt] at (1.00,0.86) {$t(u^{*}){=}t(v)$};

  \draw[thick] (1.80,-0.40) -- ++(0,0.18);
  \node[above, fill=white, inner sep=1pt] at (1.80,-0.14) {$t(w)$};

  \draw[<->,decorate,decoration={brace,amplitude=2.5pt}]
    (1.02,-0.05) -- (1.80,-0.05)
    node[midway, below=2pt, fill=white, inner sep=1pt] {$E_w\!\sim\!\mathrm{Exp}(N(w))$};
\end{tikzpicture}
\caption{\textbf{Exact-mode race.} First arrival $t(v)$; winner reuses $t(v)$; each non-winner uses offset propagation: $t(w)=t(v)+E_w$ with independent $E_w\!\sim\!\mathrm{Exp}(N(w))$.}
\label{fig:dag-right}
\end{subfigure}

\caption{\textbf{Race-coupled search overview.} (a) Context indexing enforces the child-partition property. (b) Exponential race at a node: lazy offset propagation justifies winner reuse and independent residuals for others (Lemma~\ref{lem:child-indep}).}
\label{fig:dag-race}
\endgroup
\end{figure}

\section{Child Minima, Independence, and Offset Propagation}
Let $\mathrm{Ch}(v)=\{u_1,\ldots,u_k\}$. For each child $u$, define
\[
t(u):=\min_{P\in\mathcal P(u)} E_P \ \sim\ \Exp\!\bigl(N(u)\bigr),
\]
so that $t(v):=\min_{u\in\mathrm{Ch}(v)} t(u)$ and
\[
I_v:=\argmin_{u\in\mathrm{Ch}(v)} t(u).
\]

\begin{lemma}[Child minima and residual independence]\label{lem:child-indep}
Under Assumption~\ref{as:local-finiteness} and the child-partition property, condition on $t(v)=T$ and $I_v=u^{*}$. Then for each $w\neq u^{*}$,
\[
t(w)\ \overset{\mathrm{d}}{=}\ T + E_w,\qquad E_w \sim \Exp\!\bigl(N(w)\bigr),
\]
and the residuals $\{E_w\}_{w\ne u^{*}}$ are mutually independent and independent of $T$.
\end{lemma}

\emph{Sketch.} Because leaves under distinct children are disjoint and leaf-level $E_P$ are independent, the child minima $\{t(u)\}_{u\in\mathrm{Ch}(v)}$ are independent with $t(u)\sim\Exp(N(u))$. Conditioning on $t(v)=T$ and $I_v=u^{*}$ implies $t(w)>T$ for all $w\ne u^{*}$. By memorylessness,
\[
t(w)-T\mid t(w)>T \ \sim\ \Exp\!\bigl(N(w)\bigr),
\]
independently across $w$ and independent of $T$. Equivalently, superposing the child-level Poisson point processes and applying Palm disintegration (Slivnyak) shows that deleting the atom at $T$ leaves independent Poisson processes on $(T,\infty)$, yielding the same conclusion. In \textbf{Exact mode}, this justifies \emph{lazy offset propagation}: the winner child reuses $t(u^{*})=t(v)$, while each non-winner draws an independent residual $E_w\sim\Exp(N(w))$ on first touch.

\section{Counts Policy, Safe Surrogates, and $\kappa$ Protocol}\label{sec:surrogate}
Let $N(v)=|\mathcal P(v)|$ be the exact leaf count. When exact counts are unavailable, we use \emph{upper bounds} $\Nub(v)\!\ge\!N(v)$ together with a surrogate race $\hat t(\cdot)$ and \emph{disable winner reuse}. If $\Nub(v)=0$ then $\mathcal P(v)=\emptyset$ and $v$ is pruned immediately; otherwise $\Nub(v)\ge1$.

\paragraph{Lazy surrogate instantiation and anchoring.}
In Surrogate mode we instantiate $\hat t$ \emph{on pop} except at the \emph{root} (on push):
\begin{itemize}
\item \textbf{Root:} draw $U(\mathrm{root})\!\in\!(0,1)$ and set $\hat t(\mathrm{root})=-\log(1-U(\mathrm{root}))/\Nub(\mathrm{root})$; push with $\Keyhat(\mathrm{root})=M_\tau(\mathrm{root})-\log\hat t(\mathrm{root})$.
\item \textbf{Internal $v$:} when $v$ is popped, draw $U(v)$ and set $\hat t(v)=-\log(1-U(v))/\Nub(v)$. For each child $u$, push the \emph{provisional} key $M_\tau(u)-\log\hat t(v)$. \textbf{All} child keys remain parent-anchored (provisional) until the child is popped.
\end{itemize}
(Optional: a monotone variant logs residual uniforms and sets $\hat t(u)=\hat t(v)+\mathrm{Exp}(\Nub(u))$; not required.)

\begin{proposition}[Upper-bound rates are safe]\label{prop:ubsafe}
If $\Nub(\cdot)\!\ge\!N(\cdot)$ for all nodes, there is a monotone inverse-CDF coupling (use the \emph{same} node-level uniform $U(v)$) with
\[
\hat t(v)=\frac{-\log(1-U(v))}{\Nub(v)}\ \le\ \frac{-\log(1-U(v))}{N(v)}\ =:\ t^{\circ}(v)\ \stackrel{d}{=}\ t(v).
\]
Under the canonical quantile coupling for Exp\((N(v))\) (using the same \(U(v)\)), identify \(t(v)\equiv t^{\circ}(v)\).
Hence \( -\log \hat t(v)\!\ge\!-\log t(v)\) and \(\Keyhat(v)\!\ge\!\Key(v)\) (sound but potentially looser).
\end{proposition}

\noindent\textit{$\kappa$ tightening.} When the validator can recover $N(v)$ for some nodes, it applies $\kappa(v)=\log\!\big(N(v)/\Nub(v)\big)\!\le\!0$ to tighten keys (otherwise $\kappa=0$), without changing the frontier invariant.

\paragraph{Quantile-coupled correction $\kappa$.}
With $U(v)\sim\mathrm{Unif}(0,1)$ and using the \emph{same} $U(v)$ for both rates (node-level quantile coupling),
\[
\hat t(v)=\frac{-\log(1-U(v))}{\Nub(v)},\quad 
t(v)=\frac{-\log(1-U(v))}{N(v)},\quad
-\log \hat t(v)+\kappa(v)= -\log t(v),
\]
where $\kappa(v):=\log\!\big(N(v)/\Nub(v)\big)$. 
\emph{Protocol.} The \textbf{router} logs $(U(v),\Nub(v))$. The \textbf{validator}, when it can compute $N(v)$ from public counts (\textsc{SuffixCountDP}), derives $\kappa(v)$ and tightens the key; otherwise no correction is applied and the surrogate key remains conservative.

\section{Keys and Best-First Search: Two-Mode Theory}\label{sec:bestfirst}
\begin{lemma}[Provisional-key safety]\label{lem:provisional}
For a child $u$ of $v$, $t(u)\ge t(v)$ a.s.\ in the true race. Hence $M_\tau(u)-\log t(v)$ upper-bounds the realized subtree maximum under $u$.
\end{lemma}
\emph{Justification.} Since $t(u)\!\ge\!t(v)$, we have $-\log t(v)\!\ge\!-\log t(u)$; and by the standard PaM/A*-style bound, the realized maximum under $u$ is $\le M_\tau(u)-\log t(u)$, so the parent-anchored provisional key is an upper bound as well.

\begin{theorem}[Frontier coverage: Exact mode]\label{thm:frontier-exact}
Assuming admissible, race-independent $M_\tau$, child partition, and local finiteness, best-first search with keys $\Key(v)=M_\tau(v)-\log t(v)$ has the following property. Let $\mathcal{F}$ be the PQ frontier and define $B^{*}:=\max_{\text{expanded } P}\bigl(s_{\mathrm{det}}(P)-\log E_P\bigr)$. Then \emph{every unexpanded leaf lies under some $v\in\mathcal{F}$ whose key upper-bounds the realized subtree maximum}. $\max_{P\in\mathcal{P}(v)}\bigl(s_{\mathrm{det}}(P)-\log E_P\bigr)$. Consequently, halting when $\max_{v\in\mathcal{F}}\Key(v)\le B^{*}$ is stop-correct (no unexpanded leaf can improve $B^{*}$). Ties occur with probability $0$ for real uniforms; with fixed-point uniforms they are resolved via the logged lexicographic rule.
\end{theorem}

\begin{lemma}[Surrogate frontier coverage]\label{lem:frontier-surrogate}
If only $\Nub$ are available, compute \emph{provisional} child keys as $\Keyhat(u)=M_\tau(u)-\log \hat t(\mathrm{parent}(u))$ (\emph{no winner reuse}). Then for all frontier nodes $v$,
\[
\max_{P\in\mathcal P(v)} \bigl(s_{\mathrm{det}}(P)-\log E_P\bigr) \;\le\; \Keyhat(v)\quad\text{a.s.}
\]
\emph{Proof idea.} Since $\hat t(v)\le t(v)$ a.s.\ (Prop.~\ref{prop:ubsafe}) and $t(u)\ge t(v)$ a.s.\ (Lemma~\ref{lem:provisional}), we have $-\log \hat t(v)\ge -\log t(v)\ge -\log t(u)$. Thus $\Keyhat(u)=M_\tau(u)-\log \hat t(v) \ge M_\tau(u)-\log t(u)$, which upper-bounds the realized subtree maximum under $u$.
\end{lemma}

\begin{theorem}[Best-first: two modes]\label{thm:two-modes}
In Exact mode, finalized child keys are $M_\tau(u)-\log t(u)$; in Surrogate mode, all child keys remain parent-anchored \emph{provisional} values $M_\tau(u)-\log \hat t(\mathrm{parent}(u))$ until $u$ is popped. In both modes, the frontier invariant holds and the corresponding stop rule is sound for the realized run.
\end{theorem}

\paragraph{Algorithm sketch (high level).}
Initialize the root according to mode (Exact: $t(\mathrm{root})$ via $U$ and $N$; Surrogate: $\hat t(\mathrm{root})$ via $U$ and $\Nub$; Fallback: bound only). Maintain a PQ $\mathcal F$ of nodes keyed by $\Key$ or $\Keyhat$, and an incumbent $B^{*}$. While true: (i) if $\max_{v\in\mathcal F}\Key(v)\!\le\!B^{*}$ (\textbf{Exact}) or $\max_{v\in\mathcal F}\Keyhat(v)\!\le\!B^{*}$ (\textbf{Surrogate}), halt with a run-wise certificate; (ii) pop $v$. If $v$ is a leaf $P$, compute $V(P)=s_{\mathrm{det}}(P)-\log E_P$ (from the logged or PRF uniform per mode) and update $B^{*}\leftarrow\max(B^{*},V(P))$. Else expand children: \textbf{Exact} reuses $t(v)$ for the winner and sets $t(w){=}t(v){+}E_w$ for others; \textbf{Surrogate} anchors all child keys at $-\log\hat t(v)$ until each child is popped. Ties use the logged lexicographic rule; uniforms and budgets are logged for replay. Full pseudocode is in Appx.~M (Alg.~\ref{alg:bestfirst}).

\section{Admissible $M_\tau$: Boxed Recipe and Guidance}\label{sec:mtau}
\paragraph{Score units and signs.}
All logs are natural. \emph{$M_\tau$ upper-bounds only the deterministic part of the score reachable from $v$} (the race noise is handled by $-\log t$ in the key). Let $d(v)$ be a certified upper bound on steps to a leaf. Let $c^s_{\max}\!\ge\!0$ upper-bound the \emph{per-step increase} in the deterministic score used by $M_\tau$ (R1), and let $c^s_{\min}\!>\!0$ lower-bound the \emph{per-step decrease} in the best-attainable remainder (for truncation).

\begin{boxedremark}
\textbf{Recipe R1 (per-step envelope).} If at most $d(v)$ steps remain and each step increases the deterministic score by $\le c^s_{\max}$, then
\[
M_\tau(v)= s_{\mathrm{det}}(\text{prefix}(v)) + d(v)\,c^s_{\max}
\]
is admissible.

\textbf{Recipe R2 (monotone remainder).} If a monotone $\psi$ satisfies $s_{\mathrm{det}}(\text{prefix}\circ r)\le \psi(\text{prefix})$ for any remainder $r$, then

\[
M_\tau(v)=\psi(\text{prefix}(v))
\]
is admissible.

\textbf{Choosing R1 vs.\ R2.} Prefer R2 when a tight monotone envelope exists; otherwise use R1 with certified $d(v)$ and $c^s_{\max}$. Both recipes are race-independent.
\end{boxedremark}

\paragraph{LSE truncation (pointer).}
An absolute LogSumExp decomposition certifying tail truncation (with proof) is given in Appx.~\ref{app:scale-trunc}, Eq.~\eqref{eq:lse-absolute}.

\section{Acyclicity Certificate and Guardrails}
\label{sec:guards}
\emph{Moved to Appendix~\ref{sec:guards-app}.} We verify termination via a potential $\Phi$ that decreases by at least $\eta>0$ per expansion and downgrade on violations; a replayable log and examples appear in Appx.~\ref{sec:guards-app}.

\section{From Shared-Node DAGs to Context-Indexed Prefix--DAGs}
\label{sec:compiler-transform-stub}
\emph{Moved to Appendix~\ref{app:compiler-transform}.} We context-index nodes (state, prefix context) to enforce child partition; certificates and transform details appear in Appx.~\ref{app:compiler-transform}.

\section{Fallback: PRF-per-Leaf (NoCert) and Work Bound}\label{sec:fallback}
When counts fail, we switch to PRF-per-leaf: each leaf $P$ has a deterministic uniform $U_P=\mathrm{PRF}(\text{id}(P))\in(0,1)$ (open-interval mapping). We reuse the \emph{same} $U_P$ for both perturbations: $G(P)=-\log(-\log U_P)$ and $E_P=-\log(1-U_P)$. The (heuristic) PaM value we rank by is
\[
V_{\text{pam}}(P)\ :=\ \frac{G(P)}{\tau}-C(P)\;-\;\log E_P
\qquad\text{(marked \texttt{NoCert}).}
\]

\paragraph{Claim type and stop rule.}
\textbf{Fallback is \emph{NoCert}}: its stop rule is \emph{heuristic} for ranking/latency (not a run-wise proof). Maintain a second PQ $\mathcal L$ of \emph{materialized leaves} ordered by realized augmented value; PQ ties follow the same lexicographic rule as $\mathcal F$. Internal-node expansions are governed by admissible $M_\tau$-based bounds; when an internal node $v$ is popped, enumerate children and materialize any leaves into $\mathcal L$. Stop when no tracked bound suggests a leaf can exceed the incumbent $B^{*}$; the ledger marks \texttt{claim\_type=NoCert}.

\begin{lemma}[Fallback work bound]\label{lem:fallback}
Let $\mathcal A$ be the best-first that pops leaves in non-increasing realized augmented value (with fixed PRF uniforms) and stops once every remaining frontier node’s exact-leaf LSE upper bound is $<B^{*}$. Then PRF fallback enumerates exactly the leaves $\mathcal A$ would enumerate, plus at most one additional deferred-sibling evaluation per expanded internal node.
\end{lemma}
\emph{Proof sketch.} The PRF fixes a total order on leaf perturbations, so realized values are deterministic. Node-wise upper bounds match the baseline’s pruning logic; materializations occur exactly when the baseline would pop those leaves, with a one-per-node overhead due to deferred siblings. Full proof and accounting appear in Appx.~E.

\section{Numerics, Logging, and Validator}
\label{sec:numerics}
Fixed-point layouts (Q0.64 for $U$, Q64.64 for $-\log t$/$-\log\hat t$), tie handling, ledger schema, and deterministic replay checks are given in Appendix~\ref{app:ledger-schema}. We retain the open-interval mapping $U=(x+0.5)\,2^{-64}$, record the PRF salt/domain in the ledger for replay, and use compensated transforms (e.g., \texttt{log1p}) for $\log(1-U)$.

\section{Privacy Posture and Post-Processing}
\label{sec:privacy}
All routing randomness (uniforms, PRG seeds, HKDF labels) and counts/surrogates are fixed independently of the raw input $X$; grammar and allow-listed operations are public (pre-LDP). By the \emph{post-processing theorem} for (L)DP \citep{DworkRoth2014}, any (possibly adaptive) function of DP outputs and public data preserves DP. With modern accounting choices such as GDP \citep{Dong2022GDP}, Fourier/PLD \citep{ZhuDongWang2022FourierAccountant,Ghazi2022EvolvingPLD}, and the discrete Gaussian mechanism \citep{Canonne2020DiscreteGaussian}, the ledger—being a function of public grammar/config, logged $U$, and public or privatized counts—preserves privacy. If $N$ depends on user context, validator-side $\kappa$ must be computed from \emph{public} counts only; otherwise $\kappa=\perp$ and surrogate keys remain conservative. Seeds are derived via HKDF \citep{RFC5869}; Rényi composition follows \citet{Mironov2017RDP}. \textbf{Threat-model caveat.} If grammar selection or model/adapter choice depends on \emph{non-privatized} raw input, the run-wise correctness claim still holds, but the DP posture must exclude that branch (router marks \texttt{claim\_type=NoCert} or declines DP claims).

\section{Privacy-Budgeted Multi-LLM (Summary)}\label{sec:budgeted-routing}
We attach a per-request budget controller (privacy, price, latency) that \emph{post-processes} DP outputs and public data only, keeping $M_\tau$ admissible and race-independent and thus preserving the two-mode invariant and stopping rules. Operationally, smaller frontier slack $\Key(\cdot){-}B^{*}$ (or $\Keyhat(\cdot){-}B^{*}$) justifies spending to tighten bounds. \emph{Full interface, accountant, selection rule, and proofs of budget safety/soundness are in Appx.~\ref{app:budgeted}}. 
Empirically, Appendix~\ref{app:adapter-table} confirms that inference $\varepsilon$ remains $0.0$ on average across all adapters (post-processing), while Table~\ref{tab:adapter-summary} reports which trained adapters (with certified $(\varepsilon_{\text{train}},\delta_{\text{train}})$) were actually selected in our runs.

\section{Experiments: Illustrative Evidence and Replay}
All runs are \emph{Mac/commodity-hardware reproducible} with fixed seeds and full ledgers. We report three families: (i) comparisons to baselines on two synthetic suites; (ii) tightness, stress, and overhead diagnostics; and (iii) an adversarial case where distribution-level pruning is \emph{unsound} but our run-wise certificate holds. \textbf{Appendix~\ref{app:real-pipeline}} adds a small real tool-use pipeline (retrieval$\to$summarize$\to$calculator) with ledger replay and validator $\kappa$‑tightening.

\paragraph{Suites.}
\textbf{Suite A (balanced trees):} depth $D\!\in\!\{3,4,5\}$, branching $B=3$.
\textbf{Suite B (random partition DAGs):} layers $L\!\in\!\{3,4\}$, branching $B=3$; generated with the child-partition property (post compilation to a prefix--DAG).
A small toolchain DAG (retrieval$\to$summary$\to$calc) is used in sanity checks and the ledger section (cf.\ contemporary tool-use agents \citep{Schick2023Toolformer,Yao2023ReAct,Patil2024Gorilla}).

\paragraph{Baselines.}
(1) \textbf{NoCert greedy-by-bound:} expand $\arg\max_v M_\tau(v)$ (no run-wise certificate). 
(2) \textbf{Beam($K$):} standard beam search of width $K$ (no certificate).
(3) \textbf{Dist-level:} prune using $\mathbb{E}[-\log \Emin]=\gamma+\log N$ for $\Emin\!\sim\!\mathrm{Exp}(N)$; ignores realized coupling and is \emph{not} a per-run certificate.

\paragraph{Tightness and stress diagnostics.}
Figure~\ref{fig:tightness} plots the empirical CDF of $\Key(v)-\mathrm{RSM}(v)$ over frontier nodes at stop (where $\mathrm{RSM}(v)=\max_{P\in\mathcal P(v)}(s_{\mathrm{det}}(P)-\log E_P)$; cf.\ Appx.~A): both modes concentrate near $0$; \textbf{Exact} is tight at the stopping frontier (slack $\le 0$, often $0$). Figure~\ref{fig:nub} sweeps the Surrogate $\Nub$ factor and reports expansions vs.\ the validator’s $\kappa$-tightening potential (fraction with strict $\kappa=\log(N/\Nub)\!<\!0$, i.e., $\Nub>N$): larger $\Nub$ induces more conservative routing (more expansions) but yields more strict-tightening opportunities.

\paragraph{Metrics.}
We report node expansions, wall-clock, stopping slack, and validator replay time; full per-run ledgers and seeds are provided for deterministic replay (Appx.~\ref{app:exp-details}; ledger schema/protocol in Appx.~\ref{app:ledger-schema}).

\paragraph{Real tool-use (summary; full table in Appx.~\ref{app:real-pipeline}).}
On $n{=}20$ queries, the pipeline exhibits the expected two-mode ordering: \textbf{Exact} $<$ \textbf{Surrogate} $<$ \textbf{Fallback} in expansions (mean per query $6.20$, $11.60$, $16.60$; 95\% CIs $1.12$, $0.81$, $0.81$), with wall-time per query $1.16$, $2.75$, and $1.20$\,ms, respectively. \emph{Ledger replay} passes deterministically for all runs, and \emph{Surrogate} shows validator $\kappa$‑tightening on all nodes (232/232; mean $\kappa=-0.549$). Stop-slack at termination is $0$ (tight stop by construction).

\paragraph{Adapter transparency (summary; full table in Appx.~\ref{app:adapter-table}).}
We expose which DP-trained adapters the controller considered and which were actually chosen, together with their training budgets $(\varepsilon_{\text{train}},\delta_{\text{train}})$ and the average \emph{inference} $\varepsilon$ consumed (should be 0 under post-processing). Table~\ref{tab:adapter-summary} summarizes the adapters used in our Mac-only runs; the complete per-mode breakdown appears in Appendix~\ref{app:adapter-table} and is programmatically generated by \texttt{make adapter-table}.

\begin{table}[t]
\centering
\small
{
    \setlength{\tabcolsep}{6pt}
    \begin{tabular}{lcccc}
    \toprule
    Adapter (tier) & $(\varepsilon_{\text{train}},\delta_{\text{train}})$ & Selected (runs) & Selected in mode & Inference $\varepsilon$ used \\
    \midrule
    LoRA-A-small (Small)   & $(2.0,\ 10^{-6})$ & 240 & Exact      & 0.0 \\
    LoRA-B-medium (Medium) & $(3.5,\ 10^{-6})$ & 680 & Surrogate  & 0.0 \\
    (no-adapter)           & ---               & 402 & Fallback   & 0.0 \\
    \bottomrule
    \end{tabular}
}
\caption{\textbf{Adapter transparency (main-text summary).} Counts reflect our reproducibility runs (Mac), and will vary with seeds/session; the full per-mode table with all adapters is in Appx.~\ref{app:adapter-table}. Under our router’s post-processing posture, inference $\varepsilon$ remains zero.}
\label{tab:adapter-summary}
\end{table}

\paragraph{Ledger overhead and validator replay.}
The ledger records uniforms, counts (and $\Nub$), $\kappa$ (validator), keys (raw/tight), budgets, and guard flags in fixed point (formats in Appx.~\ref{app:ledger-schema}); detailed overheads (bytes, appends, parse/validate time) are reported in Appx.~\ref{app:ledger-overhead}, Table~\ref{tab:overhead}.

\begin{figure}[!t]
\centering
\begin{subfigure}[t]{0.49\linewidth}
  \centering
  \includegraphics[width=\linewidth]{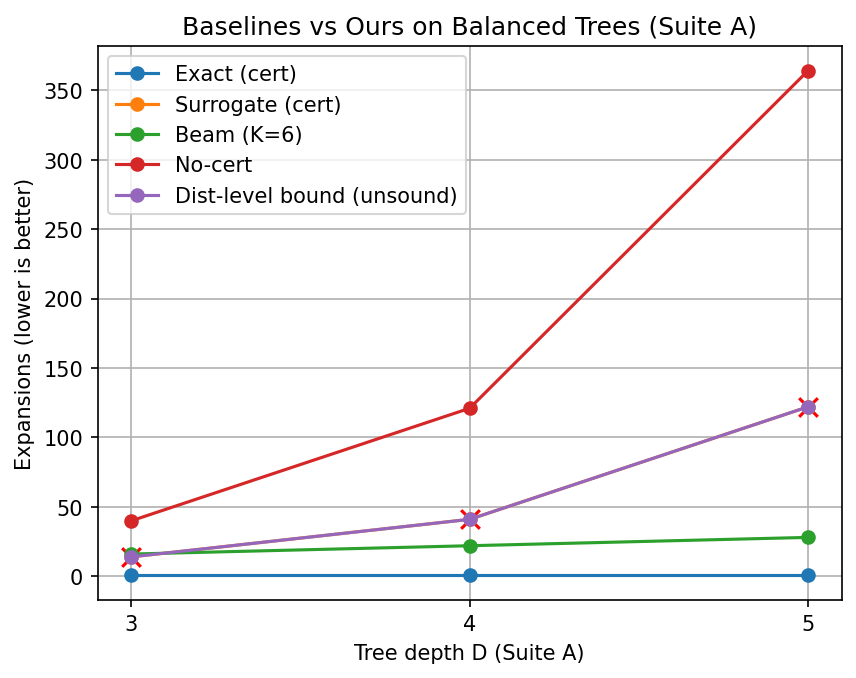}
  \caption{Suite A (balanced).}
  \label{fig:baselinesA}
\end{subfigure}\hfill
\begin{subfigure}[t]{0.49\linewidth}
  \centering
  \includegraphics[width=\linewidth]{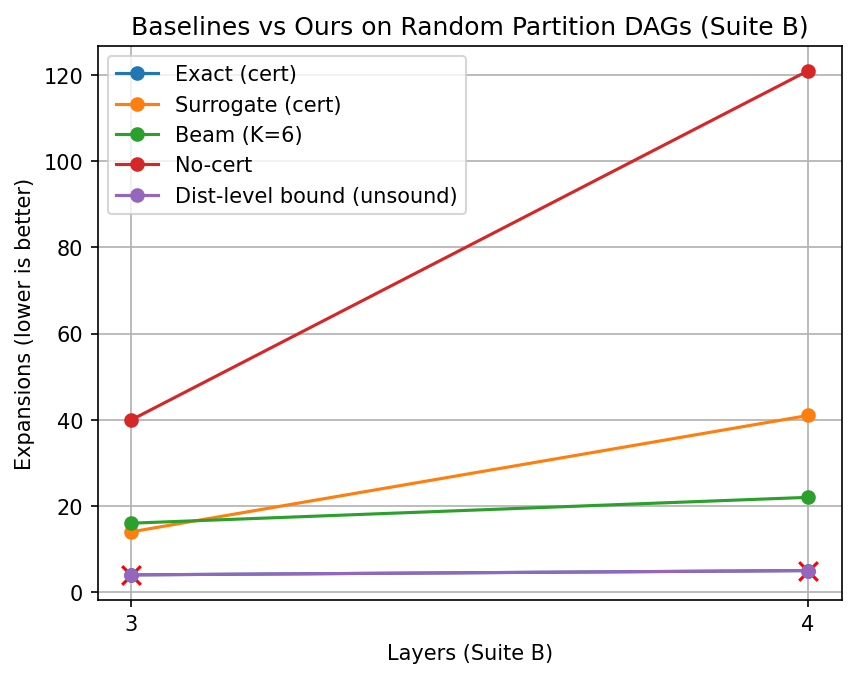}
  \caption{Suite B (random).}
  \label{fig:baselinesB}
\end{subfigure}

\begin{subfigure}[t]{0.49\linewidth}
  \centering
  \includegraphics[width=\linewidth]{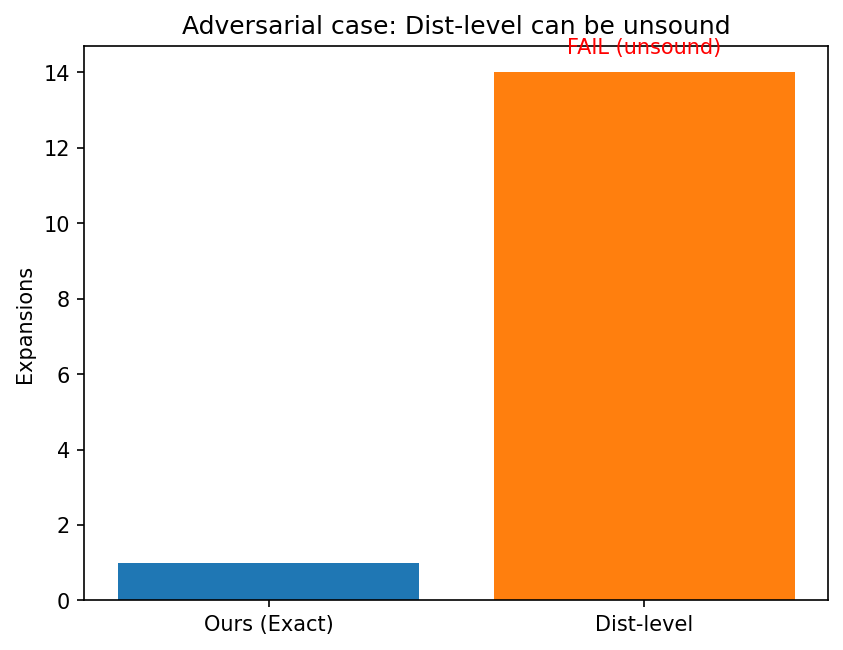}
  \caption{Adversarial fail of distribution-level bound.}
  \label{fig:adversarial}
\end{subfigure}
\caption{\textbf{Baselines across two suites + adversarial.}
Node expansions (lower is better) on (a) balanced trees and (b) random partition DAGs (after compilation to a prefix--DAG).
On these toy graphs, \textbf{Exact} expands least; \textbf{Surrogate} is slightly larger but certified.
\textbf{Beam} and \textbf{Greedy-by-bound} expand more, and the distribution-level heuristic using $\gamma{+}\log N$ can be \emph{unsound} (marked $\times$ when it prunes the realized winner).
(c) In an adversarial DAG, the distribution-level heuristic prunes the realized winner; our run-wise certificate does not.}
\label{fig:baselinesAB}
\end{figure}

\paragraph{Adversarial case against distribution-level pruning.}
In Fig.~\ref{fig:adversarial}, the Dist-level baseline prematurely prunes the branch containing the true realized winner because it reasons with $\mathbb{E}[-\log \Emin]$ (e.g., $\gamma{+}\log N$) rather than the \emph{realized} race. Our run-wise certificate is sound: if we stop, every unexpanded leaf is provably dominated by a frontier key computed under the same realized race, so the true winner is never pruned early.

\paragraph{Budgeted controller (system add-on).}
The controller is orthogonal to the certificate. In our Mac toy, a strict SLO collapses the frontier to a single operating point (Appx.~\ref{app:pareto}); with a looser SLO the same code yields multiple Pareto points. We include these for completeness; full ablations and Pareto plots are in Appx.~\ref{app:pareto}.

\emph{Scaling and truncation diagnostics appear in Appx.~\ref{app:scale-trunc}.}

\subsubsection*{LLM Usage (Disclosure)}
A general-purpose LLM assisted with non-novel utilities (code/plot scaffolding, LaTeX formatting, copy-editing). Authors take full responsibility for all content.

\section{Related Work}
\textbf{A* Sampling}~\citep{MaddisonTarlowMinka2014} and \textbf{Gumbel processes / exponential races}~\citep{Kingman1993,Resnick1992,Maddison2016PoissonMC,Huijben2021GumbelReview} provide \emph{distribution-level} guarantees; see also planning with Gumbel improvements~\citep{Danihelka2022GumbelPlanning}. \textbf{Gumbel-Top-$k$}~\citep{Kool2019} targets ranked sampling, not frontier-sound early stopping. \textbf{Perturb-and-MAP} for MRFs~\citep{PapandreouYuille2011,HazanJaakkola2012} leverages perturbations for optimization but does not couple pruning keys to the \emph{same} randomness that realizes the run. Classical A* admissibility~\citep{HartNilsRaphael1968,Pearl1984} informs our $M_\tau$ recipes but lacks stochastic coupling and ledger/validator design. On privacy, modern accountants (GDP, Fourier/PLD, discrete Gaussian) complement our post-processing posture~\citep{DworkRoth2014,Dong2022GDP,ZhuDongWang2022FourierAccountant,Ghazi2022EvolvingPLD,Canonne2020DiscreteGaussian}; recent work on DP fine-tuning for LMs provides systems context~\citep{Yu2022DPFineTuneLM,Charles2024UserLevelDP,Chua2024MindPrivacyUnit}. Tool-use agents~\citep{Schick2023Toolformer,Yao2023ReAct,Patil2024Gorilla} motivate budget-aware routing under latency/cost constraints. \textbf{Our novelty} is a \emph{single-race, run-wise} coupling with a \emph{two-mode frontier invariant}, plus an auditable ledger consistent with LDP post-processing.

\section{Conclusion}
We present a theory-first, audit-ready early-stopping method for PaM routing that certifies \emph{the realized run} on context-indexed prefix--DAGs. A \emph{two-mode} frontier invariant (Exact vs.\ Surrogate) closes a gap in race-based methods; the ledger/validator protocol makes deployment practical under LDP post-processing; a budgeted multi-LLM controller (with DP-trained LoRA adapters; see \citep{Hu2021LoRA}) adds system utility while preserving run-wise soundness. \emph{Scope:} guarantees assume child-partition, local finiteness, and admissible, race-independent $M_\tau$; on violation we downgrade to \texttt{NoCert}.
\FloatBarrier   
\clearpage      

\ifarxiv
\subsubsection*{Reproducibility Statement}
All code and scripts to reproduce our results are provided as the ancillary file
\texttt{code.zip} attached to this arXiv submission (see the “Ancillary files” section on the abstract page). 
The ZIP includes \texttt{README.md} with exact commands, environment pins 
(\texttt{requirements.txt}/\texttt{environment.yml}), fixed seeds, and dataset checksums.
\fi

\subsubsection*{Ethics Statement}
This work uses only public data and synthetic workloads; no human subjects or private user data were collected.

\ifarxiv
\bibliographystyle{abbrvnat}
\fi
\bibliography{iclr2026_conference}

\appendix

\section{Appendix A: Child-Minima Independence (Poisson/Palm)}
\label{app:child-indep}

\subsection*{Terminology (Formal)}
\begin{definition}[Realized subtree maximum]
For node $v$ under a fixed realized exponential race $\{E_P\}$,
\[
\mathrm{RSM}(v) := \max_{P\in \mathcal P(v)}\big(s_{\mathrm{det}}(P) - \log E_P\big).
\]
\end{definition}

\begin{definition}[Run-wise certificate]
We say the algorithm returns a \emph{run-wise certificate} if, when it stops with
\[
\max_{u\in \mathcal F}\Key(u) \le B^{*} \quad \text{(frontier $\mathcal F$)},
\]
it follows that for every unexpanded leaf $P$ we have $s_{\mathrm{det}}(P)-\log E_P \le B^{*}$ almost surely under the \emph{same} realized race.
\end{definition}

\noindent\emph{Measurability note.} The lexicographic tie resolver (on public IDs) is logged and replayed by the validator, so the stop rule is a measurable function of the ledger.

\paragraph{Full proof.}
Model each child $u$ as a superposition of $N(u)$ i.i.d.\ Poisson point processes of rate $1$. Superpose across children to a rate-$N(v)$ process on $\mathbb R_+$. Condition on the first arrival at time $T$ belonging to child $u^{*}$. By Palm calculus (Slivnyak), removing the atom at $T$ leaves independent Poisson processes unchanged on $(T,\infty)$, giving independent $\mathrm{Exp}(N(w))$ residuals for $w\ne u^{*}$, independent of $T$ \citep{Kingman1993,Resnick1992}.

\section{Appendix B: Frontier Coverage (Exact and Surrogate Modes)}
\label{app:frontier}
\textbf{Invariant (restated).} Fix the realized race $\{E_P\}$ (Appx.~A). At any time, every unexpanded leaf lies under some frontier node $v\in\mathcal F$ whose key upper-bounds the realized subtree maximum:
\[
\begin{aligned}
\mathrm{RSM}(v)
  &:= \max_{P\in\mathcal{P}(v)}\bigl(s_{\mathrm{det}}(P)-\log E_P\bigr),\\
\mathrm{RSM}(v)
  &\le
  \begin{cases}
    \Key(v)   & \text{(Exact)},\\
    \Keyhat(v)& \text{(Surrogate)}.
  \end{cases}
\end{aligned}
\]
\noindent\textit{where }\(\Key(v)=M_\tau(v)-\log t(v)\) and \(\Keyhat(v)=M_\tau(v)-\log \hat t(\mathrm{parent}(v))\).

Here and below, in Surrogate mode $\Keyhat(\cdot)$ denotes the PQ key:
for the root, $\Keyhat(\mathrm{root})=M_\tau(\mathrm{root})-\log\hat t(\mathrm{root})$;
for any not-yet-popped internal node $u$,
$\Keyhat(u)=M_\tau(u)-\log\hat t(\mathrm{parent}(u))$.

\textbf{Base.} After pushing $\mathrm{root}$, admissibility of $M_\tau$ implies $\mathrm{RSM}(\mathrm{root})\le M_\tau(\mathrm{root})-\log t(\mathrm{root})$ in Exact, and $\le M_\tau(\mathrm{root})-\log\hat t(\mathrm{root})$ in Surrogate.

\textbf{Step (Exact).} Pop $v$; by Lemma~\ref{lem:child-indep}, $t(u^{*}){=}t(v)$ for the winner and $t(w){=}t(v){+}E_w$ with $E_w\!\sim\!\Exp(N(w))$ independent otherwise. Admissibility gives $\mathrm{RSM}(u)\le M_\tau(u)-\log t(u)$ for each child; pushing children preserves the invariant.

\textbf{Step (Surrogate).} Pop $v$; instantiate $\hat t(v)$ and push each child $u$ with \emph{parent-anchored} provisional key $M_\tau(u)-\log\hat t(v)$. By Proposition~\ref{prop:ubsafe} and Lemma~\ref{lem:provisional}, $-\log\hat t(v)\ge -\log t(v)\ge -\log t(u)$ almost surely, hence $M_\tau(u)-\log\hat t(v)\ge M_\tau(u)-\log t(u)\ge \mathrm{RSM}(u)$, preserving the invariant. Ties are null a.s.\ with real $U$; with fixed-point $U$ they are resolved via the logged lexicographic rule (measurable and replayable).

\section{Appendix C: $M_\tau$ Soundness (R1/R2) and Prefix-Monotonicity}
\paragraph{Formal statements.}
\textbf{R1 (per-step envelope).} If at most $d(v)$ steps remain and each step can increase the deterministic part of $s$ by at most $c^s_{\max}$, then
\[
M_\tau(v)=s(\mathrm{prefix}(v)) + d(v)\,c^s_{\max}
\]
is admissible: for any completion $r$, $s(\mathrm{prefix}(v)\circ r)\le M_\tau(v)$.

\textbf{R2 (monotone remainder).} If a monotone $\psi$ satisfies $s(\mathrm{prefix}\circ r)\le \psi(\mathrm{prefix})$ for all remainders $r$, then $M_\tau(v)=\psi(\mathrm{prefix}(v))$ is admissible.

\paragraph{Proof sketches.}
R1 is a telescoping bound over the remaining $d(v)$ steps. R2 follows by the definition of $\psi$. In both cases $M_\tau$ depends only on grammar/prefix constraints and public caps (e.g., max depth), not on stochastic tool/model outputs; hence it is race-independent and remains admissible under any model choice.

\paragraph{Optional prefix-monotonicity.}
If $\psi$ is nonincreasing under prefix extension (tighter with longer prefixes), then $M_\tau$ inherits a useful monotonicity that can reduce PQ churn; this is not required for soundness.

\section{Appendix D: Unsafe Lower-Bound Counterexample}
Lower-bound rates $N_{\mathrm{lb}}(\cdot)\le N(\cdot)$ can \emph{under-estimate} keys and break frontier soundness. A concrete numeric instance:

\paragraph{Two-leaf toy.}
Let $v$ have two children, each with one leaf: $N(v)=2$ and (for simplicity) $M_\tau(v)=0$. Draw $U=0.5$. Then in the true race
\[
t(v)=\frac{-\log(1-U)}{N(v)}=\frac{-\log(0.5)}{2}\approx 0.3466,\qquad -\log t(v)\approx 1.0609.
\]
If a (wrong) lower bound $N_{\mathrm{lb}}(v)=1$ is used at $v$, the surrogate gives
\[
t_{\mathrm{lb}}(v)=\frac{-\log(0.5)}{1}\approx 0.6931,\qquad -\log t_{\mathrm{lb}}(v)\approx 0.3665,
\]
so the key drops from $\;0+1.0609\;$ to $\;0+0.3665$, i.e., it \emph{underestimates} the realized subtree maximum. In this toy, with one leaf per child and $M_\tau\equiv 0$, $\mathrm{RSM}(v)=-\log\min(E_{P_1},E_{P_2})=-\log t(v)\approx 1.0609$, but the lower-bound key evaluates to $\approx 0.3665<\mathrm{RSM}(v)$, violating the invariant. This illustrates why we prohibit lower bounds and allow only \emph{upper} counts $\Nub\!\ge\!N$ (which make keys conservative).

\section{Appendix E: Fallback Work Bound (Full Proof) and Tight Example}\label{app:fallback}
We formalize the realized-score best-first baseline and give a bijection between popped leaves, plus accounting for one deferred-sibling residual per expanded internal node.

\paragraph{Baseline $\mathcal A$ (realized-score best-first).}
Fix a deterministic PRF-per-leaf $\{U_P\}$ and define heuristic PaM values $V_{\text{pam}}(P):=\frac{G(P)}{\tau}-C(P)-\log E_P$ with $E_P=-\log(1-U_P)$ and $G(P)=-\log(-\log U_P)$. Algorithm $\mathcal A$ maintains a PQ over leaves and pops in non-increasing $V_{\text{pam}}(P)$, pruning internal nodes via exact leaf-wise LSE upper bounds (as in Sec.~\ref{sec:fallback}).

\begin{theorem}[Fallback work bound (formal)]\label{thm:fallback}
Run the PRF fallback (Sec.~\ref{sec:fallback}) and the baseline $\mathcal A$ under the same PRF uniforms $\{U_P\}$ and the same admissible bounds. Then:
\begin{enumerate}
\item The sets of popped leaves satisfy
$\mathrm{Leaves}_{\text{fallback}}=\mathrm{Leaves}_{\mathcal{A}}\cup S$
with $\lvert S\rvert \le N_{\text{int}}$, where $N_{\text{int}}$ is the number of internal nodes expanded by fallback.
\item Each $P\in \mathrm{Leaves}_{\mathcal A}$ is popped by fallback in a step no later than when $\mathcal A$ pops $P$ (same realized total order), hence the incumbent $B^{*}$ matches at those times.
\end{enumerate}
\end{theorem}

\emph{Proof sketch.} (i) \textbf{Coupling.} Both procedures use the same per-leaf PRF uniforms,so $V_{\text{pam}}(P)$ are identical and totally ordered. (ii) \textbf{Pruning parity.} Internal-node upper bounds are the same, so any subtree $\mathcal A$ prunes, fallback also prunes. (iii) \textbf{Materialization parity.} Whenever $\mathcal A$ would pop a leaf $P$, fallback must have materialized $P$ (it materializes leaves when their parent is popped), so $P$ enters fallback’s leaf-PQ by that time. (iv) \textbf{Overhead.} Fallback may materialize at most one \emph{deferred sibling} per expanded internal node that $\mathcal A$ never needs to pop (because $\mathcal A$ can directly compare leaves across subtrees). Summing over expanded internal nodes yields $|S|\le N_{\text{int}}$.

\paragraph{Tight example.}
In a binary tree of depth $D{=}2$ with $M_\tau\equiv0$, arrange PRF uniforms so that along each internal node, exactly one child’s best leaf is globally uncompetitive. $\mathcal A$ never pops those leaves; fallback materializes exactly one such deferred sibling at each internal expansion, achieving the $|S|=N_{\text{int}}$ worst case.

\section{Appendix F: Acyclicity Certificate and Guardrails (moved)}
\label{sec:guards-app}
We require a potential $\Phi$ that strictly decreases at each expansion until a cap $L$.

\paragraph{Template and parameter choice.}
For pipelines with (at most) $L$ tool calls and a cost $C$ that increases by $\Delta C\ge c_{\min}>0$ per expansion, set
\[
\Phi(v)\ :=\ (L-\mathrm{depth}(v))\ +\ \alpha\,C(\mathrm{prefix}(v)).
\]
Each expansion changes $\Phi$ by
\[
\Delta\Phi\ =\ -1\ +\ \alpha\,\Delta C\ \le\ -1\ +\ \alpha\,c_{\min}.
\]
Choose $\eta\in(0,1]$ and $\alpha$ with $-1+\alpha c_{\min}\le -\eta$ (e.g., $\alpha\le (1-\eta)/c_{\min}$). Then $\Delta\Phi\le -\eta<0$, certifying termination.

\paragraph{Runtime check (deterministic).}
At each expansion, compute $\Delta\Phi$ in fixed point; if $\Delta\Phi>-\,\eta+\epsilon_{\mathrm{fp}}$ (tolerance), \emph{downgrade claim type to \texttt{NoCert}} and log \textsc{AcyclicityFail} with the node id and measured $\Delta\Phi$. A short ledger excerpt is in Appx.~\ref{app:acyc-example}.

\paragraph{Local finiteness checks.}
The compiler emits certificates for finite $N(v)$. Overflow thresholds/timeouts in \textsc{SuffixCountDP} trigger \textsc{CountFail} with node IDs; \emph{downgrade to \texttt{Surrogate}} if counts remain upper-bounded, else to \texttt{NoCert}. All transitions are logged and replayable.
\emph{Clarification.} Our \textsc{SuffixCountDP} routine operates over public grammar/compilation artifacts; when
available it returns \emph{exact} (non-private) counts. The “DP” suffix reflects its role in larger pipelines where
upstream counts may already be privatized; the validator never uses non-public counts to compute $\kappa$.

\subsection{Example and Downgrade Log}\label{app:acyc-example}
Concrete $\Phi$ for a tool-call grammar; a short ledger excerpt shows \textsc{AcyclicityFail} and claim-type downgrade.

\begin{verbatim}
{"node_id":"3f1a...e2","parent_id":"de2b...90","mode":"Exact",
 "claim_type_before":"RunWiseExact","guards":[],
 "phi_before":"12.0000","phi_after":"11.2000","delta_phi":"-0.8000",
 "eta":"1.0000"}

{"node_id":"7a98...bd","parent_id":"3f1a...e2","mode":"Exact",
 "claim_type_before":"RunWiseExact","guards":["AcyclicityFail"],
 "phi_before":"11.2000","phi_after":"10.5000","delta_phi":"-0.7000",
 "eta":"1.0000","claim_type_after":"NoCert","budget_event":"None"}
\end{verbatim}

\section{Appendix G: Ledger Parser and Validator Checklist}
\label{app:ledger-parser}
\textbf{Fixed-point layouts.} Log $U$ as Q0.64 (open interval via $U=(x+0.5)\,2^{-64}$ with $x\in\{0,\ldots,2^{64}-1\}$), $-\log t$ and $-\log\hat t$ as Q64.64, and $\kappa$ as Q32.32 (validator-computed). Compute $\log(1-U)$ using \texttt{log1p(-U)}; if unavailable, use compensated transforms \citep{Kahan1965}. Record \textsc{NumClamp} if any value would overflow the target format.

\textbf{Ties.} With fixed-point $U$, equal draws have small but non-zero probability; log a \texttt{tie\_token} (child permutation index) and resolve lexicographically on public IDs. The validator replays the same rule, ensuring measurability.

\paragraph{Ledger/validator semantics (clarified).}
The ledger carries \texttt{privacy\_scope=post\_processing\_only} and a transparency field \texttt{router\_rdp\_eps} (with selected order $\alpha$). These reflect router-internal RDP bookkeeping for per-edge noises logged as metadata; \emph{they are not a spend}. In our formulation the router is pure post-processing of upstream LDP; the authoritative per-request budget is the LDP composition. We leave \texttt{eps\_delta.eps} \emph{unset} unless LDP atoms are present and validated. The reference validator recomputes \texttt{router\_rdp\_eps} for auditability and separately checks any LDP atoms if provided.

\textbf{LangGraph integration.}
Our reference implementation expresses the controller, accountant, and tool calls as \emph{LangGraph} nodes \citep{LangGraph2024}; for tool APIs and cross-runtime handshakes, the Model Context Protocol (MCP) offers a compatible interface \citep{MCP2024Anthropic}. Ledger appends occur at node boundaries and replay deterministically.

\textbf{Ledger schema.} We log IDs, keys, budgets, guards, and DP metadata; the full field list is in Appendix~\ref{app:ledger-schema}.

\textbf{Parser \& determinism.} Ledger is NDJSON \citep{NDJSON}; numbers are decimal strings parsed into fixed point with overflow checks and IEEE-754 round-to-nearest-even. Determinism verified across \{clang/libm, MSVCRT\} using identical fixed-point arithmetic and locale-independent parsing. A \(\sim\)100-LoC reference parser (pseudo) and validator checklist: recomputation, $\kappa$ tightening, stop-rule verification, guardrail- and budget-induced downgrades; \emph{verify Eq.~\eqref{eq:budget-score} inequalities}, \emph{recompute $(\varepsilon,\delta)$ from atoms}, and \emph{check adapter metadata presence/consistency} (\texttt{adapter\_id}, \texttt{dp\_cert\_id}, $(\varepsilon_{\text{train}},\delta_{\text{train}})$). Also record PRF domain/salt to make $U_P=\mathrm{PRF}(\text{id}(P))$ replayable.

\subsection{Full Ledger Schema}\label{app:ledger-schema}
\begin{table}[t]
\centering
\begin{tabular}{@{}p{0.34\linewidth}p{0.62\linewidth}@{}}
\toprule
\textbf{Field} & \textbf{Type / Q-format} \\ \midrule
node\_id, parent\_id & 128-bit UUIDv7 (time-ordered UUID; cf.\ \citep{RFC4122}) \\
mode, claim\_type & enum \{Exact/Surrogate/Fallback\}, \{RunWiseExact, TruncationOnly, NoCert\} \\
privacy\_scope & enum \{post\_processing\_only, \ldots\} \\
U, Nub, kappa & Q0.64, 64-bit unsigned; Q32.32 (validator-computed) \\
key\_raw, key\_tight & Q64.64 (router/validator) \\
router\_rdp\_eps, alpha\_selected & Q32.32, integer (router-internal transparency) \\
eps\_delta.eps, eps\_delta.delta & \emph{unset unless LDP atoms present}, decimal string for $\delta$ \\
price\_spent, price\_cap & uint64 (cents) \\
sla\_ms & uint32 (milliseconds) \\
budget\_event & enum \{None, BudgetFail\} \\
dkey\_pred, dkey\_real & Q32.32 (predicted vs.\ realized slack change) \\
model\_id & string (selected model/tool identifier) \\
adapter\_id & string (hash of weights+config: LoRA $\alpha$, rank, target modules) \\
dp\_cert\_id & string (certificate identifier for adapter training) \\
eps\_train, delta\_train & Q32.32, decimal string (adapter training DP params) \\
tie\_token & uint32 (child perm index) \\
guards & bitset \{CountFail, AcyclicityFail, NumClamp, Timeout, BudgetFail, CapExceeded\} \\
\bottomrule
\end{tabular}
\end{table}

\subsection{Ledger Overhead Details}\label{app:ledger-overhead}
\begin{table}[t]
\centering
\caption{\textbf{Ledger overhead and validator replay (deterministic).} Bytes on disk, approximate append operations, parser latency, and validator recomputation time. Overheads are small (sub-ms parse/validate for these toy runs) and scale roughly linearly with node events.}
\label{tab:overhead}
{
    \setlength{\tabcolsep}{5pt}%
    \resizebox{\linewidth}{!}{%
    \begin{tabular}{@{}lrrrr@{}}
    \toprule
    \textbf{ledger} & \textbf{bytes} & \textbf{appends} & \textbf{parse\_ms} & \textbf{validate\_ms} \\
    \midrule
    e5b1fa73-e497-4a5c-97a5-6116cea3f6e2.json & 1766  &  24 & 10.08479087613523  & 0.0639590434730053 \\
    e7525b1a-71c4-4f36-82c9-5029f574e786.json & 4672  &  68 &  9.169207885861397 & 0.08795899339020252 \\
    fc83bc2d-715d-47f1-b00b-f20f52fbf507.json & 2261  &  31 & 15.063582919538021 & 0.05320901982486248 \\
    b6d29d45-f95f-49ad-8666-167ecb14af12.json & 14790 & 203 & 11.514000128954649 & 0.17150002531707287 \\
    5d9d9188-c433-4e97-bfd3-cfc6670f8e90.json &  664  &   4 &  1.6771249938756227& 0.04050019197165966 \\
    be1ef3bc-7ee9-4e13-b078-69ae56d1cc2f.json & 14771 & 203 &  2.524791983887553 & 0.17758295871317387 \\
    428e6364-170c-4338-a884-1d1e584f6744.json & 14771 & 203 &  2.4711249861866236& 0.17779087647795677 \\
    fd51d33a-0778-43e4-ab72-02712c14fd7b.json & 14771 & 203 &  2.376958029344678 & 0.1551660243421793  \\
    a993a9c2-afcb-4408-9bdd-f81e9e41b8b3.json & 14771 & 203 &  2.3737919982522726& 0.1664168667048216  \\
    b7825d56-5a2f-4173-b3d4-0b97079e6ec3.json & 14771 & 203 &  2.2709579207003117& 0.1660841517150402  \\
    \bottomrule
    \end{tabular}%
}
}
\end{table}

\section{Appendix H: Compiler Transform Details and Partition Certificates}
\label{app:compiler-transform}
We map each node to $(\text{state},\text{prefix-context})$, where the \emph{prefix context} is a canonical serialization of the full ancestor sequence (tools/states/constraints) and “public caps” (e.g., max depth). Two parent paths that reach the same state but with different contexts produce distinct nodes, ensuring the \emph{child-partition} property after compilation to a prefix--DAG.

\paragraph{Collision handling.}
We compute a collision-resistant digest $h(\cdot)$ (e.g., SHA-256/BLAKE3) over the canonical serialization with domain separation and log \texttt{ctx\_digest} alongside a short \texttt{ctx\_repr}. The validator recomputes $h$ and checks that siblings have distinct digests; on any mismatch (or digest collision), the router downgrades to \texttt{NoCert} and logs \textsc{CountFail} with the offending node IDs. (Digest collisions are negligible in practice, but we treat any detected collision as a certification failure.)

\paragraph{Complexity.}
At search depth $D$ with max branching $B$, explored prefix contexts are $O(B^D)$. In explored regions with bounded context growth, the transform does not change asymptotic branching; otherwise we rely on guardrails (Appx.~\ref{sec:guards-app}) and make no worst-case claim.

\paragraph{Certificates emitted.}
(i) a \emph{partition certificate} that witnesses child leaf-set disjointness and $N(v)=\sum_{u\in\mathrm{Ch}(v)}N(u)$; (ii) a \emph{finiteness certificate} for local finiteness; (iii) a \emph{stable leaf-ID} policy (leaf IDs are canonical hashes of $(\text{state},\text{prefix-context},\text{leaf-spec})$) so replay is deterministic.

\section{Appendix I: Toy 4-Leaf Replay}\label{app:toy}
\paragraph{Toy 4-leaf replay.}
Root children $u_1,u_2$ with $N(u_1){=}3$, $N(u_2){=}1$; leaves under $u_1$: $P_1,P_2,P_3$; under $u_2$: $P_4$.

\emph{Logged uniforms.} $U_r{=}0.20$ (race at root), $W_r{=}0.70$ (winner selection at root), $U_{u_2}{=}0.37$ (residual for $u_2$). Quantile winner: since $W_r<3/4$, $I_r{=}u_1$.

Then $t(r){=}-\log(1-0.20)/4 \approx 0.055786$, so $-\log t(r)\approx 2.884$.
Winner reuse (Exact): $t(u_1){=}t(r)$. For $u_2$, $E_{u_2}{=}-\log(1-0.37)\approx 0.462$, hence $t(u_2)\approx 0.518$ and $-\log t(u_2)\approx 0.657$.
With $M_\tau(r){=}5$, $M_\tau(u_1){=}4.5$, $M_\tau(u_2){=}4.2$:
\[
\Key(r){\approx} 5 - \log t(r) \approx 7.886,\quad
\Key(u_1){\approx} 4.5 - \log t(r) \approx 7.386,\quad
\Key(u_2){\approx} 4.2 - \log t(u_2) \approx 4.859.
\]

\emph{Surrogate mode.} With $\Nub(r){=}6$ and the same $U_r$, $\hat t(r){\approx}0.03719$ so $-\log\hat t(r){\approx}3.288$, making $\Keyhat(r){\approx}8.288$ (conservative). Children would be pushed with parent-anchored provisional keys $M_\tau(u)-\log\hat t(r)$ until popped.

\section{Appendix J: Additional System Figures}\label{app:pareto}
\begin{figure}[t]
\centering
\includegraphics[width=0.72\linewidth]{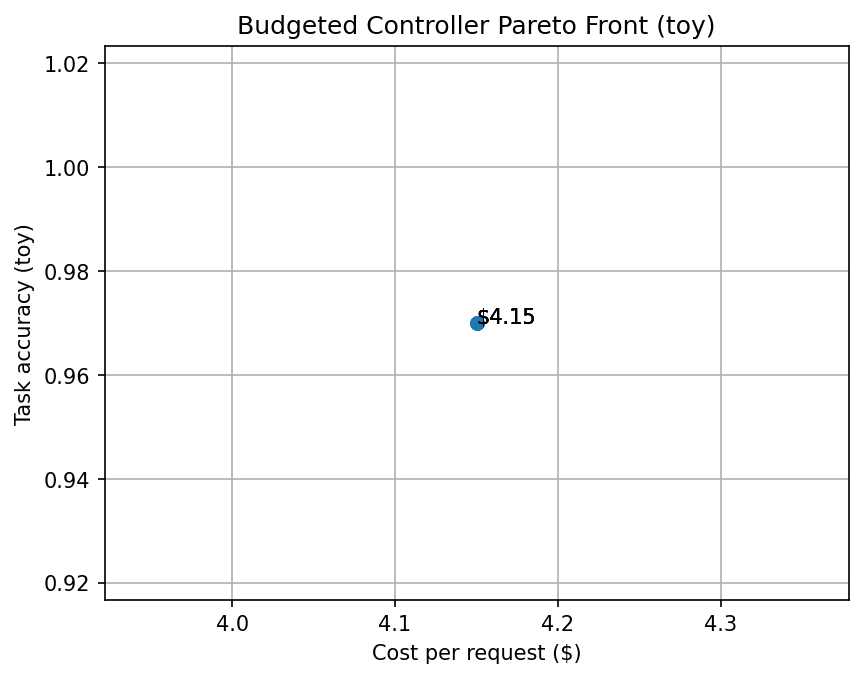}
\caption{Budgeted controller Pareto front (toy). Operating points under price/SLO; a strict SLO yields a single point.}
\end{figure}

\begin{figure}[!t]
\centering
\begin{subfigure}[t]{0.49\linewidth}
  \centering
  \includegraphics[width=\linewidth]{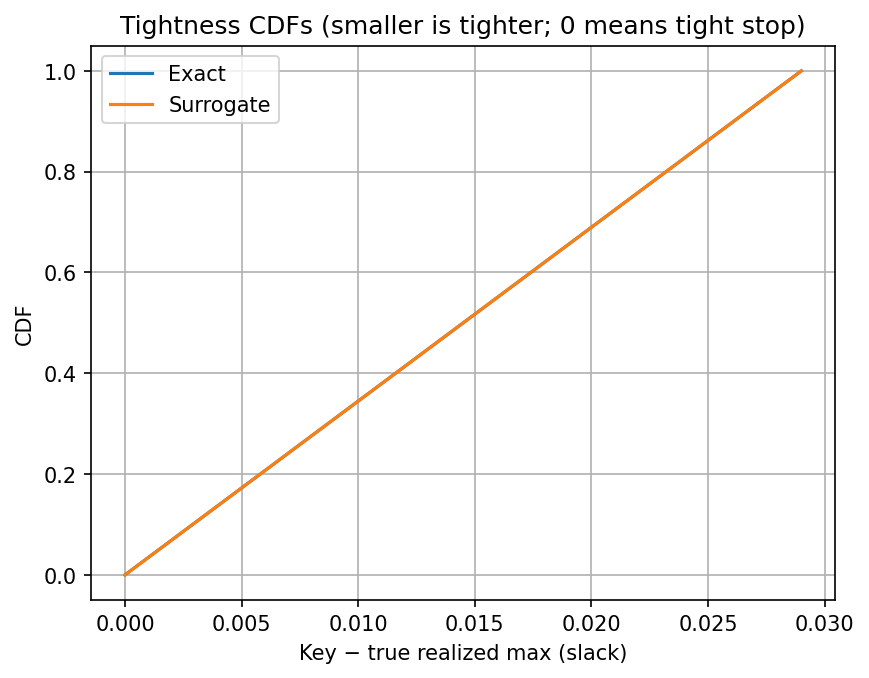}
  \caption{Tightness CDFs.}
  \label{fig:tightness}
\end{subfigure}\hfill
\begin{subfigure}[t]{0.49\linewidth}
  \centering
  \includegraphics[width=\linewidth]{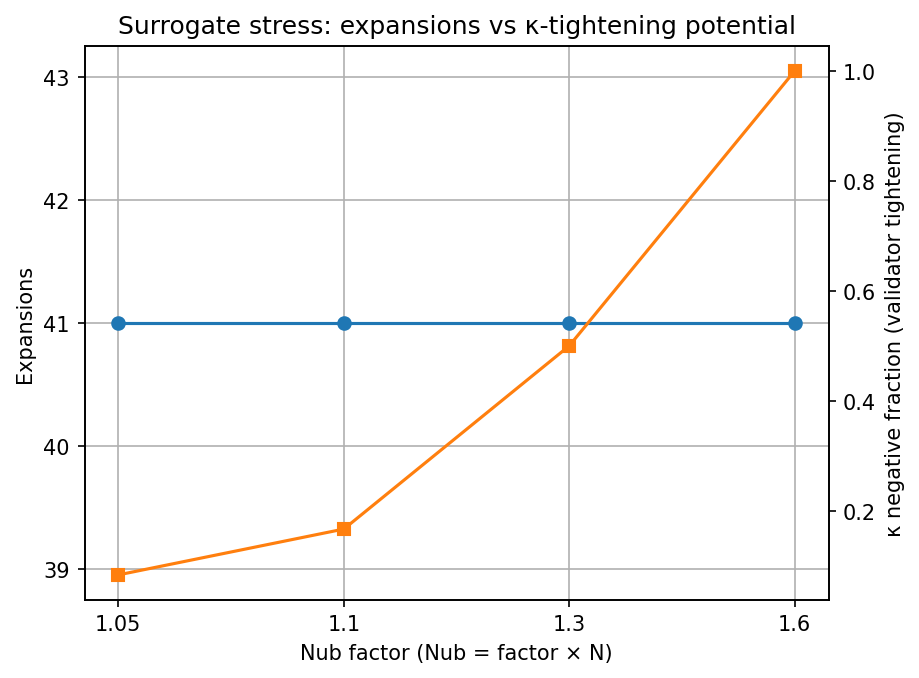}
  \caption{Surrogate stress \& $\kappa$ potential.}
  \label{fig:nub}
\end{subfigure}
\caption{\textbf{Key tightness and surrogate behavior.}
(a) CDF of $\Key-\max_{\text{subtree}}(\text{realized})$ shows \textbf{Exact} is tight at stop (slack $\approx 0$) and \textbf{Surrogate} is close because parent-anchored keys remain informative.
(b) Increasing $\Nub$ makes keys more conservative (more expansions) but yields more validator tightening (higher fraction of negative $\kappa=\log(N/\Nub)$); logs show mean $\kappa\approx -0.257$.}
\label{fig:tightness-nub}
\end{figure}

\section{Appendix K: Scaling and Truncation Diagnostics}\label{app:scale-trunc}
\subsection{Truncation in Score Units and an Absolute LogSumExp Bound}
\paragraph{Definition of $B_k$.}
For a given prefix node, let $B_k$ be an \emph{upper bound on the number of admissible leaves at depth $k$} under the grammar and compilation constraints (per-depth leaf counts after context indexing; not branching factors).

\paragraph{Absolute tail bound (no denominator).}
Let $c^s_{\min}>0$ be a per-step \emph{decrease} in the best possible score, and let $s_{\mathrm{ref}}$ be any certified upper bound on the score of any complete path under the current prefix (e.g., $s_{\mathrm{ref}}=M_\tau(\mathrm{root})$). Then the overall $\mathrm{LSE}=\log\sum_P e^{s_{\mathrm{det}}(P)}$ obeys
\begin{equation}
\mathrm{LSE}(\text{all})
\;\le\;
\max\Big(\mathrm{LSE}(\le K),\; s_{\mathrm{ref}} + \log\!\big(\sum_{k>K} B_k \, e^{-k\,c^s_{\min}}\big)\Big)
\;+\;\log 2.
\label{eq:lse-absolute}
\end{equation}
\emph{Proof.} For any $P$ at depth $k$, $s_{\mathrm{det}}(P)\le s_{\mathrm{ref}}-k c^s_{\min}$.Hence $\sum_{\text{depth}(k)} e^{s_{\mathrm{det}}(P)} \le B_k\, e^{s_{\mathrm{ref}}-k c^s_{\min}}$. Summing $k>K$ bounds the tail. Split $\sum e^{s(P)}$ into $\le K$ vs.\ $>K$ and apply $\log(x{+}y)\le \max(\log x,\log y)+\log 2$. \emph{Mental model:} if $B_k\le B^k$, the RHS tail becomes geometric with ratio $B e^{-c^s_{\min}}$.
\begin{figure}[t]
\centering
\begin{subfigure}[t]{0.49\linewidth}
  \centering
  \includegraphics[width=\linewidth]{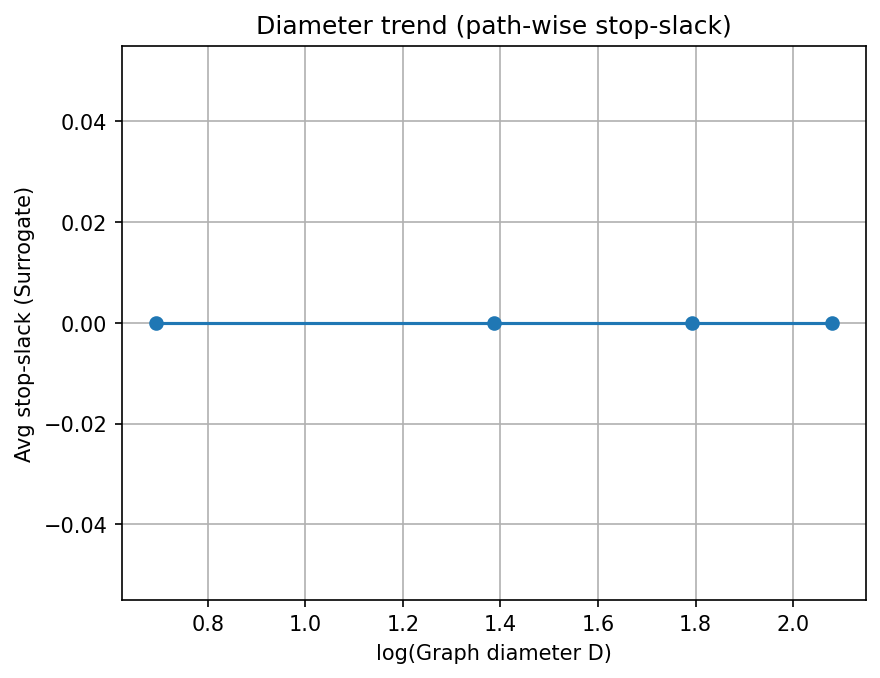}
  \caption{Diameter vs.\ stop-slack.}
  \label{fig:diameter}
\end{subfigure}\hfill
\begin{subfigure}[t]{0.49\linewidth}
  \centering
  \includegraphics[width=\linewidth]{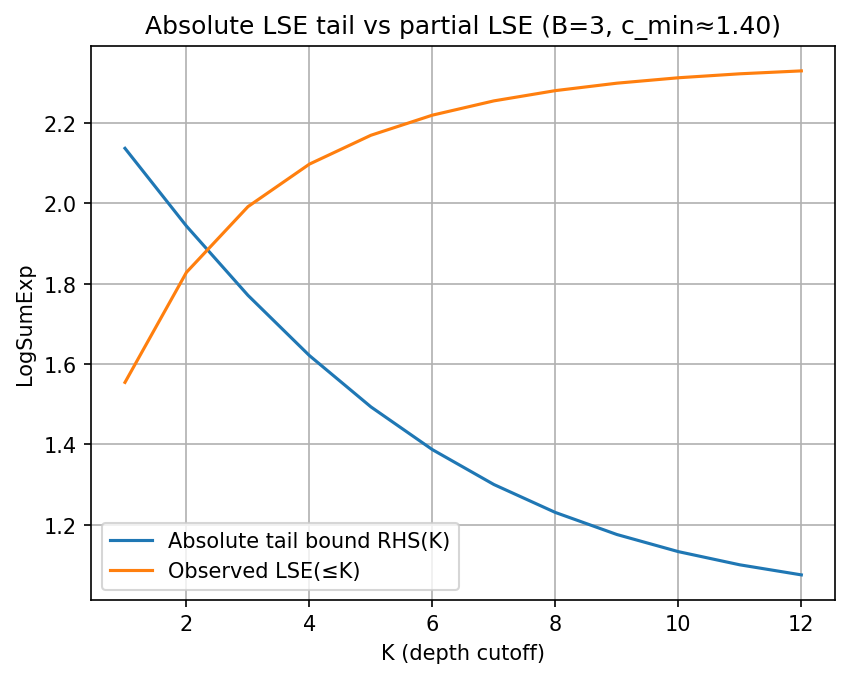}
  \caption{Absolute LSE tail vs.\ partial.}
  \label{fig:lse-tail}
\end{subfigure}
\caption{\textbf{Tight stopping and LSE truncation certificate.}
\emph{Left:} Average stop–slack $\max\{0,\,\max_{v\in\mathcal{F}}\Key(v)-B^{*}\}$ (score units) vs.\ $\log D$, where $D$ is the graph diameter (maximum root-to-leaf steps); values stay $\approx 0$, indicating tight stopping without runaway path cost as graphs grow.
\emph{Right:} Absolute LogSumExp decomposition (Eq.~\eqref{eq:lse-absolute}): the certificate uses the maximum between the partial term $\mathrm{LSE}(\le K)$ and the tail term $s_{\mathrm{ref}}+\log\!\sum_{k>K} B_k e^{-k c^s_{\min}}$ (up to the $+\log 2$ constant). When $B e^{-c^s_{\min}}<1$ the tail decays geometrically; after small $K$, the partial term dominates, certifying safe truncation in score units.}
\end{figure}

\section{Appendix L: Experimental Setup \& Reproducibility}
\label{app:exp-details}
We provide run seeds, per-run ledgers (NDJSON), and replay scripts for Mac/commodity hardware. The validator replays keys, stop rules, $\kappa$-tightening, and budget events deterministically using the fixed-point formats in Appx.~\ref{app:ledger-schema}; figures report means over seeds with identical logged uniforms and PRF domain/salt.

\section{Appendix M: Full Pseudocode}
\label{app:full-alg}

\begin{algorithm}[!t]
\caption{\textbf{Best-First PaM (Exact / Surrogate) with Fallback (NoCert)}}
\label{alg:bestfirst}
\begin{algorithmic}[1]
\State \textbf{Input:} grammar $\mathcal G$, mode $\in\{\texttt{Exact},\texttt{Surrogate},\texttt{Fallback}\}$, admissible $M_\tau$, bounds $(\{B_k\},c^s_{\min},K)$
\State Initialize frontier PQ $\mathcal F$ (key tie-breaker: lexicographic public IDs); leaf PQ $\mathcal L$ (Fallback only); $B^{*}\leftarrow-\infty$
\If{mode$==\texttt{Exact}$}
  \State compute $N(\mathrm{root})$; draw $U(\mathrm{root})\!\in\!(0,1)$; set $t(\mathrm{root})=-\log(1-U(\mathrm{root}))/N(\mathrm{root})$
  \State push $\mathrm{root}$ with $\Key(\mathrm{root})=M_\tau(\mathrm{root})-\log t(\mathrm{root})$; log $(U(\mathrm{root}),N(\mathrm{root}))$
\ElsIf{mode$==\texttt{Surrogate}$}
  \State get $\Nub(\mathrm{root})$; draw $U(\mathrm{root})\!\in\!(0,1)$; set $\hat t(\mathrm{root})=-\log(1-U(\mathrm{root}))/\Nub(\mathrm{root})$
  \State push $\mathrm{root}$ with $\Keyhat(\mathrm{root})=M_\tau(\mathrm{root})-\log \hat t(\mathrm{root})$; log $(U(\mathrm{root}),\Nub(\mathrm{root}))$
\Else \Comment{Fallback (NoCert)}
  \State push $\mathrm{root}$ with $M_\tau(\mathrm{root})$; initialize $\mathcal L\leftarrow\emptyset$; mark \texttt{claim\_type=NoCert}
\EndIf
\While{true}
  \State \textbf{(optional)} \textsc{BudgetController} updates $(\varepsilon_{\mathrm{used}}, \mathrm{price}_{\mathrm{spent}}, \mathrm{latency}_{\mathrm{acc}})$ and logs tallies
  \If{any budget exhausted} \State mode $\leftarrow$ \texttt{Fallback}; log \textsc{BudgetFail} \EndIf
  \If{mode$==\texttt{Exact}$ \textbf{and} $\max_{v\in\mathcal F}\Key(v)\le B^{*}$} \State \textbf{stop} with run-wise certificate \EndIf
  \If{mode$==\texttt{Surrogate}$ \textbf{and} $\max_{v\in\mathcal F}\Keyhat(v)\le B^{*}$} \State \textbf{stop} with run-wise certificate \EndIf
  \State $v\leftarrow\mathrm{argmax}\ \mathcal F$ \Comment{pop}
  \If{$v$ is leaf $P$}
  \If{mode$==$\texttt{Exact}}
    \State $E \leftarrow t(P)$ \Comment{leaf arrival from the realized race; no new randomness}
    \State $U_P \leftarrow 1 - e^{-E}$ \Comment{logged for replay}
  \Else \Comment{Surrogate}
    \State draw or PRF-derive $U_P\!\in\!(0,1)$;\quad $E \leftarrow -\log(1-U_P)$
  \EndIf
  \State $V \leftarrow \underbrace{(-\,C(P))}_{s_{\mathrm{det}}(P)}\;-\;\log E$
  \State $B^{*} \leftarrow \max(B^{*}, V)$;\quad \textsc{log} $U_P$;\quad \textbf{continue}
\EndIf
  \If{mode$==\texttt{Exact}$}
     \State draw $W(v)\!\in\!(0,1)$; set $I_v\leftarrow\textsc{QuantileCat}\!\big(W(v);\{N(u)/N(v)\}_{u\in\mathrm{Ch}(v)}\big)$; log $W(v)$
     \For{child $u\in\mathrm{Ch}(v)$}
       \If{$u==I_v$} \State $t(u)\leftarrow t(v)$
       \Else \State draw $E_u\sim\mathrm{Exp}(N(u))$; $t(u)\leftarrow t(v)+E_u$
       \EndIf
       \State push $u$ with $\Key(u)=M_\tau(u)-\log t(u)$; log any uniforms used
     \EndFor
  \ElsIf{mode$==\texttt{Surrogate}$}
     \State draw $U(v)$; set $\hat t(v)=-\log(1-U(v))/\Nub(v)$; log $(U(v),\Nub(v))$
     \For{child $u\in\mathrm{Ch}(v)$}
       \State push $u$ with provisional $\Keyhat(u)=M_\tau(u)-\log \hat t(v)$
     \EndFor
  \Else \Comment{Fallback}
     \For{child $u\in\mathrm{Ch}(v)$}
       \If{$u$ is leaf $P$}
         \State $U_P\leftarrow \mathrm{PRF}(\text{id}(P))$; $G\leftarrow-\log(-\log U_P)$; $\text{score}\leftarrow G/\tau - C(P)$; push $P$ into $\mathcal L$
       \Else
         \State push $u$ into $\mathcal F$ with bound $M_\tau(u)$
       \EndIf
     \EndFor
     \If{$\mathcal L\neq\emptyset$ \textbf{ and } $\max_{P\in\mathcal L}\text{score}(P)$ dominates tracked bounds}
        \State \textbf{stop} (NoCert; heuristic)
     \EndIf
  \EndIf
\EndWhile
\end{algorithmic}
\end{algorithm}

\section{Appendix N: Budgeted Controller Details}
\label{app:budgeted}
We equip the router with per-request budgets for privacy, price, and latency while preserving the two-mode frontier invariants of Secs.~\ref{sec:surrogate}--\ref{sec:bestfirst}. Let the \emph{key slack} at a frontier node $v$ be
\[
\mathrm{slack}(v)=
\begin{cases}
\Key(v)-B^{*} & \text{(Exact mode)},\\[0.15em]
\Keyhat(v)-B^{*} & \text{(Surrogate mode)}.
\end{cases}
\]
Smaller slack means we are closer to stopping; thus the \textsc{BudgetController} may (i) upgrade tools/models/adapters, or (ii) tighten bounds, \emph{provided} $M_\tau$ remains admissible and race-independent (replacing $M_\tau$ by a tighter envelope is allowed) and decisions depend only on DP outputs and public data (post-processing). These actions do not alter the two-mode frontier invariants or the mode-specific stopping rules; all budget events are logged in the ledger.

\paragraph{Controller interface.}
Each request is initialized with budgets $\big(\varepsilon_{\max},\delta\big)$ for privacy (RDP composed then converted to $(\varepsilon,\delta)$), a monetary cap $\mathrm{price}_{\max}$, and a latency SLO $\mathrm{SLO}_{\mathrm{ms}}$. The router maintains $(\varepsilon_{\mathrm{used}},\mathrm{price}_{\mathrm{spent}},\mathrm{latency}_{\mathrm{acc}})$ and logs them (Sec.~\ref{sec:numerics}). We enforce the SLO using 95th-percentile latency estimates (``P95'') with a safety factor (e.g., $1.2\times$); violations log \textsc{BudgetFail} and trigger downgrade (to cheaper tools/models or \texttt{Fallback} if needed).

\paragraph{RDP accountant (upstream LDP only).}
Upstream LDP emits atoms $(\alpha_i,\varepsilon_i(\alpha_i))$ (stub and hop noises). The router composes them online via Rényi DP, $\varepsilon_{\mathrm{tot}}(\alpha)=\sum_i \varepsilon_i(\alpha)$, and converts to $(\varepsilon,\delta)$ at validation time \citep{Mironov2017RDP,DworkRoth2014}. Routing randomness and keys are post-processing of LDP outputs and public grammar (Sec.~\ref{sec:privacy}); therefore they do not consume privacy budget.

\begin{proposition}[Budget safety under post-processing]\label{prop:budget-postproc}
Let $\mathcal M$ be the upstream LDP mechanism and $\mathcal R$ the router whose randomness/decisions are measurable w.r.t.\ $\mathcal M(X)$ and public artifacts. Then for every order $\alpha>1$,
\[
D_\alpha\!\big(\mathcal R\!\circ\!\mathcal M(X)\,\big\|\,\mathcal R\!\circ\!\mathcal M(X')\big)\ \le\ D_\alpha\!\big(\mathcal M(X)\,\big\|\,\mathcal M(X')\big),
\]
so $\mathcal R\!\circ\!\mathcal M$ satisfies the same RDP curve as $\mathcal M$ and does not increase $(\varepsilon,\delta)$ at any target $\delta\!\in\!(0,1)$.
\end{proposition}
\emph{Proof sketch.} RDP is closed under post-processing \citep{DworkRoth2014,Mironov2017RDP}. Given the measurability condition, the router’s randomness (uniforms, tie tokens) and decisions are functions of $\mathcal M(X)$ and public inputs, hence cannot increase privacy loss. \emph{Note:} if the controller adds any \emph{new} DP mechanism, its atoms must be included in $\varepsilon_{\mathrm{tot}}(\alpha)$.

\paragraph{Budget-aware model/hop selection (multi-LLM).}
At an expansion of $v$, choose a tool/model $m\in\mathcal M$ with attributes $\{\mathrm{price}_m,\mathrm{latency}_m,\varepsilon_m\}$ and an estimated \emph{key-slack reduction} $\widehat{\Delta\Key}(v,m)\ge0$. Pick
\begin{equation}
\label{eq:budget-score}
m^\ast \in \operatorname*{arg\,max}_{m\in\mathcal M}
\ \frac{\widehat{\Delta\Key}(v,m)}
{\alpha\,\mathrm{price}_m + \beta\,\mathrm{latency}_m + \gamma\,\varepsilon_m}
\quad
\text{s.t.}\quad
\begin{aligned}
&\varepsilon_{\mathrm{used}}+\varepsilon_m \le \varepsilon_{\max},\\
&\mathrm{price}_{\mathrm{spent}}+\mathrm{price}_m \le \mathrm{price}_{\max},\\
&\mathrm{latency}_{\mathrm{acc}}+\mathrm{latency}_m \le \mathrm{SLO}_{\mathrm{ms}},
\end{aligned}
\end{equation}
with tradeoff weights $\alpha,\beta\!>\!0$ and $\gamma\!\ge\!0$ (to put the denominator on a common scale). Set $\gamma\!=\!0$ when downstream hops add \emph{no} additional LDP noise (i.e., $\varepsilon_m\!=\!0$ unless a hop injects new privatization). Model/tool selection itself is post-processing and free, provided $M_\tau$ remains admissible and race-independent (tightening $M_\tau$ is allowed). This controller is orthogonal to tool-use methods but complementary in practice \citep{Schick2023Toolformer,Yao2023ReAct,Patil2024Gorilla}. When adapters or fine-tuning are employed, recent DP results guide budget constraints \citep{Yu2022DPFineTuneLM,Charles2024UserLevelDP,Chua2024MindPrivacyUnit}.

\paragraph{Calibration.}
Optionally wrap $\widehat{\Delta\Key}$ with a conformal upper bound to get a conservative gain $\widehat{\Delta\Key}^{\uparrow}$; this affects only selection, not key validity. The ledger logs predicted vs.\ realized slack changes as \texttt{dkey\_pred} and \texttt{dkey\_real}.

\paragraph{Soundness w.r.t.\ keys and stop rules.}
Budgeted selection does not modify the race $\{t(\cdot),\hat t(\cdot)\}$ nor the admissibility of $M_\tau$; it only changes which actions we execute to \emph{compute/tighten} deterministic bounds. Hence the frontier coverage results and Theorem~\ref{thm:two-modes} remain valid.

\begin{proposition}[Frontier soundness under budgeted selection]\label{prop:budget-sound}
Suppose $M_\tau$ remains admissible and race-independent, and the controller obeys the constraints in \eqref{eq:budget-score}. Then (i) the frontier invariant of Lemma~\ref{lem:frontier-surrogate} (and Theorem~\ref{thm:frontier-exact} in Exact mode) holds unchanged; (ii) when budgets are exhausted and we downgrade to \texttt{NoCert}, the run remains replayable and the ledger/validator protocol remains correct. Moreover, if $M_\tau$ depends only on grammar/prefix constraints (not on stochastic tool outputs), it remains admissible under any model choice.
\end{proposition}
\emph{Proof idea.} Selection changes neither $t(\cdot)$ nor the parent anchoring of $\hat t(\cdot)$; keys are still computed as in Sec.~\ref{sec:bestfirst}. Downgrade alters only the claim type and stop rule (Algorithm~\ref{alg:bestfirst}, Sec.~\ref{sec:fallback}); replayability follows from logging uniforms/budgets/tie tokens.

\paragraph{Validator responsibilities.}
Validators recompute the RDP composition from logged atoms and re-derive $(\varepsilon,\delta)$; confirm that the inequalities in \eqref{eq:budget-score} hold at every step; verify any \textsc{BudgetFail}-induced downgrade; and check that any applied $\kappa$ comes from public counts only. Because budgets and keys are logged in fixed point (Sec.~\ref{sec:numerics}), these checks are deterministic.

\section{Appendix O: Real Tool-Use Pipeline (Small Eval)}
\label{app:real-pipeline}

We compile a retrieval$\to$summarize$\to$calculator pipeline into a context-indexed prefix--DAG whose children partition descendant leaves; the search runs in Exact/Surrogate/Fallback modes with a run-wise ledger. We evaluate $n{=}20$ queries on commodity Mac hardware. Table~\ref{tab:real-pipeline} reports expansions, wall-clock, stop-slack at termination, ledger replay, and surrogate $\kappa$‑tightening (from public counts). Repro: \verb|make real-pipeline| (writes \texttt{appendix\_table\_real\_pipeline.csv}).

\begin{table}[t]
\centering
\footnotesize
{
    \setlength{\tabcolsep}{3pt}
    \resizebox{\linewidth}{!}{%
    \begin{tabular}{l r r r l l l l}
    \toprule
    Mode & Expansions & Wall (s) & Stop-slack & Ledger & Exp/q (±95\%) & Wall/q ms (±95\%) & \shortstack{$\kappa$-tighten\\(mean)} \\
    \midrule
    Exact     & 124 & 0.023 & 0.000 & pass & $6.20\pm1.12$  & $1.16\pm0.64$ & n/a \\
    Surrogate & 232 & 0.055 & 0.000 & pass & $11.60\pm0.81$ & $2.75\pm2.34$ & yes (232/232), $\overline{\kappa}=-0.549$ \\
    Fallback  & 332 & 0.024 & 0.000\footnotemark & pass & $16.60\pm0.81$ & $1.20\pm0.20$ & n/a \\
    \bottomrule
    \end{tabular}%
    }
} 
\caption{\textbf{Real tool-use micro‑experiment (retrieval$\to$summarize$\to$calculator), $n{=}20$ queries.}
We compare three routing modes: \textbf{Exact} (run‑wise certificate with exact counts), \textbf{Surrogate} (safe upper‑bound counts; keys conservative but validator tightens with $\kappa=\log(N/\Nub)$), and \textbf{Fallback} (no run‑wise certificate).
\emph{Expansions} (lower is better) and \emph{Wall} report total cost; per‑query means $\pm$95\% CIs are shown; \emph{Stop‑slack} $=0$ indicates a tight stop (\,$\max_{v\in\mathcal F}\mathrm{key}(v)\le B^{*}$\,); \emph{Ledger} shows deterministic replay; \emph{$\kappa$‑tighten} reports nodes tightened and mean $\kappa$ (more negative is tighter).
Exact expands least; Surrogate expands more but is ex‑post tightened everywhere; Fallback explores the most.}

\label{tab:real-pipeline}
\end{table}

\footnotetext{Fallback is NoCert; we report $0$ for comparability, but stop-slack is not a run-wise guarantee in this mode.}

\section{Appendix P: DP-Trained Adapters (Transparency Table)}
\label{app:adapter-table}

We list the adapters considered by the controller, their training DP certificates $(\varepsilon_{\text{train}},\delta_{\text{train}})$, and how often they were selected per mode. We also report the mean \emph{inference} $\varepsilon$ used, which remains zero under our post-processing posture. The CSV is generated by \verb|make adapter-table| (file: \texttt{appendix\_table\_adapters.csv}) and can be re-created on any Mac with our repo.

\begin{table}[t]
\centering
\footnotesize
\caption{\textbf{DP‑trained adapters and whether the controller used them (transparency).}
Each row lists an adapter considered by the router, its DP training certificate
$(\varepsilon_{\text{train}},\delta_{\text{train}})$, how many times the budget‑aware controller selected it
in each routing mode (Exact/Surrogate/Fallback) and in total, and the average per‑hop inference privacy
$\varepsilon$ actually consumed. Under our post‑processing posture, inference $\varepsilon{=}0$ indicates no
additional DP noise at inference; training $\varepsilon$ is incurred offline. Counts are aggregated over our
reproducibility runs and will vary with seeds.}
\label{tab:adapter-full}
{
    \setlength{\tabcolsep}{5pt}
    \resizebox{\linewidth}{!}{%
    \begin{tabular}{@{}l l l c c r r r r c@{}}
    \toprule
    \multicolumn{1}{c}{Adapter} &
    \multicolumn{1}{c}{Tier} &
    \multicolumn{1}{c}{DP cert} &
    \multicolumn{1}{c}{$\varepsilon_{\text{train}}$} &
    \multicolumn{1}{c}{$\delta_{\text{train}}$} &
    \multicolumn{4}{c}{Chosen (runs)} &
    \multicolumn{1}{c}{Inference $\varepsilon$ (avg)} \\
    \cmidrule(lr){6-9}
     &  &  &  &  & Exact & Surrogate & Fallback & Total &  \\
    \midrule
    LoRA-A-small   & Small  & certA & 2.0 & $1{\times}10^{-6}$ & 240 & 0   & 0   & 240 & 0.0 \\
    LoRA-B-medium  & Medium & certB & 3.5 & $1{\times}10^{-6}$ & 0   & 680 & 0   & 680 & 0.0 \\
    LoRA-C-large   & Large  & certC & 6.0 & $1{\times}10^{-6}$ & 0   & 0   & 0   & 0   & 0.0 \\
    (no-adapter)   & None   & ---   & --- & ---               & 0   & 0   & 402 & 402 & 0.0 \\
    \bottomrule
    \end{tabular}}
}
\end{table}

\end{document}

%% file: math_commands.tex
\usepackage{amsmath,amsfonts,bm}

\def\eqref#1{equation~\ref{#1}}

\def\1{\bm{1}}

\DeclareMathAlphabet{\mathsfit}{\encodingdefault}{\sfdefault}{m}{sl}
\SetMathAlphabet{\mathsfit}{bold}{\encodingdefault}{\sfdefault}{bx}{n}

\DeclareMathOperator*{\argmin}{arg\,min}

%% file: iclr2026_conference.bbl
\begin{thebibliography}{33}
\providecommand{\natexlab}[1]{#1}
\providecommand{\url}[1]{\texttt{#1}}
\expandafter\ifx\csname urlstyle\endcsname\relax
  \providecommand{\doi}[1]{doi: #1}\else
  \providecommand{\doi}{doi: \begingroup \urlstyle{rm}\Url}\fi

\bibitem[NDJ()]{NDJSON}
Ndjson: Newline delimited json.
\newblock \url{https://ndjson.org/}.

\bibitem[Lan(2024)]{LangGraph2024}
Langgraph: A stateful orchestration framework for agentic applications.
\newblock \url{https://langchain-ai.github.io/langgraph/}, 2024.

\bibitem[MCP(2024)]{MCP2024Anthropic}
Introducing the model context protocol (mcp).
\newblock \url{https://www.anthropic.com/news/model-context-protocol}, November 2024.

\bibitem[Canonne et~al.(2020)Canonne, Kamath, and Steinke]{Canonne2020DiscreteGaussian}
C.~L. Canonne, G.~Kamath, and T.~Steinke.
\newblock The discrete gaussian for differential privacy.
\newblock In \emph{Advances in Neural Information Processing Systems (NeurIPS)}, 2020.
\newblock URL \url{https://papers.neurips.cc/paper_files/paper/2020/file/b53b3a3d6ab90ce0268229151c9bde11-Paper.pdf}.

\bibitem[Charles et~al.(2024)Charles, Ganesh, McKenna, McMahan, Mitchell, Pillutla, and Rush]{Charles2024UserLevelDP}
Z.~Charles, A.~Ganesh, R.~McKenna, H.~B. McMahan, N.~Mitchell, K.~Pillutla, and K.~Rush.
\newblock Fine-tuning large language models with user-level differential privacy.
\newblock \emph{arXiv preprint arXiv:2407.07737}, 2024.

\bibitem[Chua et~al.(2024)Chua, Ghazi, Huang, Kamath, Kumar, Liu, Manurangsi, Sinha, and Zhang]{Chua2024MindPrivacyUnit}
L.~Chua, B.~Ghazi, Y.~Huang, P.~Kamath, R.~Kumar, D.~Liu, P.~Manurangsi, A.~Sinha, and C.~Zhang.
\newblock Mind the privacy unit! user-level differential privacy for language model fine-tuning.
\newblock \emph{arXiv preprint arXiv:2406.14322}, 2024.

\bibitem[Danihelka et~al.(2022)Danihelka, Guez, Schrittwieser, and Silver]{Danihelka2022GumbelPlanning}
I.~Danihelka, A.~Guez, J.~Schrittwieser, and D.~Silver.
\newblock Policy improvement by planning with gumbel.
\newblock In \emph{International Conference on Learning Representations (ICLR)}, 2022.
\newblock URL \url{https://openreview.net/pdf?id=bERaNdoegnO}.

\bibitem[Dong et~al.(2022)Dong, Roth, and Su]{Dong2022GDP}
J.~Dong, A.~Roth, and W.~J. Su.
\newblock Gaussian differential privacy.
\newblock \emph{Journal of the Royal Statistical Society: Series B}, 84\penalty0 (1):\penalty0 3--37, 2022.
\newblock \doi{10.1111/rssb.12454}.

\bibitem[Dwork and Roth(2014)]{DworkRoth2014}
C.~Dwork and A.~Roth.
\newblock \emph{The Algorithmic Foundations of Differential Privacy}, volume~9 of \emph{Foundations and Trends in Theoretical Computer Science}.
\newblock Now Publishers Inc., 2014.

\bibitem[Ghazi et~al.(2022)Ghazi, Kamath, Kumar, and Manurangsi]{Ghazi2022EvolvingPLD}
B.~Ghazi, P.~Kamath, R.~Kumar, and P.~Manurangsi.
\newblock Faster privacy accounting via evolving discretization.
\newblock In \emph{Proceedings of the 39th International Conference on Machine Learning (ICML)}, volume 162 of \emph{Proceedings of Machine Learning Research}, pages 7470--7483, 2022.
\newblock URL \url{https://proceedings.mlr.press/v162/ghazi22a/ghazi22a.pdf}.

\bibitem[Hart et~al.(1968)Hart, Nilsson, and Raphael]{HartNilsRaphael1968}
P.~E. Hart, N.~J. Nilsson, and B.~Raphael.
\newblock A formal basis for the heuristic determination of minimum cost paths.
\newblock \emph{IEEE Transactions on Systems Science and Cybernetics}, 4\penalty0 (2):\penalty0 100--107, 1968.

\bibitem[Hazan and Jaakkola(2012)]{HazanJaakkola2012}
T.~Hazan and T.~Jaakkola.
\newblock On the partition function and random maximum a-posteriori perturbations.
\newblock In \emph{Proceedings of the 29th International Conference on Machine Learning (ICML)}, 2012.

\bibitem[Hu et~al.(2022)Hu, Shen, Wallis, Allen-Zhu, Li, Wang, Wang, and Chen]{Hu2021LoRA}
E.~J. Hu, Y.~Shen, P.~Wallis, Z.~Allen-Zhu, Y.~Li, S.~Wang, L.~Wang, and W.~Chen.
\newblock Lora: Low-rank adaptation of large language models.
\newblock In \emph{International Conference on Learning Representations (ICLR)}, 2022.
\newblock arXiv:2106.09685.

\bibitem[Huijben et~al.(2023)Huijben, Kool, Paulus, and van Sloun]{Huijben2021GumbelReview}
I.~A.~M. Huijben, W.~Kool, M.~B. Paulus, and R.~J.~G. van Sloun.
\newblock A review of the gumbel-max trick and its extensions for discrete stochasticity in machine learning.
\newblock \emph{IEEE Transactions on Pattern Analysis and Machine Intelligence}, 45\penalty0 (2):\penalty0 1353--1371, 2023.
\newblock \doi{10.1109/TPAMI.2022.3157042}.

\bibitem[Kahan(1965)]{Kahan1965}
W.~M. Kahan.
\newblock Pracniques: Further remarks on reducing truncation errors.
\newblock \emph{Communications of the ACM}, 8\penalty0 (1):\penalty0 40, 1965.
\newblock \doi{10.1145/363707.363723}.

\bibitem[Kingman(1993)]{Kingman1993}
J.~F.~C. Kingman.
\newblock \emph{Poisson Processes}, volume~3 of \emph{Oxford Studies in Probability}.
\newblock Clarendon Press, Oxford, 1993.

\bibitem[Kool et~al.(2019)Kool, van Hoof, and Welling]{Kool2019}
W.~Kool, H.~van Hoof, and M.~Welling.
\newblock Stochastic beams and where to find them: The {G}umbel-top-$k$ trick for sampling sequences without replacement.
\newblock In \emph{Proceedings of the 36th International Conference on Machine Learning}, volume~97 of \emph{Proceedings of Machine Learning Research}, pages 3499--3508. PMLR, 2019.
\newblock URL \url{https://proceedings.mlr.press/v97/kool19a.html}.

\bibitem[Krawczyk and Eronen(2010)]{RFC5869}
H.~Krawczyk and P.~Eronen.
\newblock Hmac-based extract-and-expand key derivation function (hkdf).
\newblock RFC 5869, 2010.
\newblock URL \url{https://www.rfc-editor.org/rfc/rfc5869}.

\bibitem[Leach et~al.(2005)Leach, Mealling, and Salz]{RFC4122}
P.~J. Leach, M.~Mealling, and R.~Salz.
\newblock A universally unique identifier (uuid) urn namespace.
\newblock RFC 4122, 2005.
\newblock URL \url{https://www.rfc-editor.org/rfc/rfc4122}.

\bibitem[Liu et~al.(2025)Liu, Zhu, Zha, Gao, Zhong, White, and Qiu]{Liu2025DPLoRA_TMIS}
X.-Y. Liu, R.~Zhu, D.~Zha, J.~Gao, S.~Zhong, M.~White, and M.~Qiu.
\newblock Differentially private low-rank adaptation of large language models using federated learning.
\newblock \emph{ACM Transactions on Management Information Systems}, 16\penalty0 (2):\penalty0 1--24, 2025.
\newblock \doi{10.1145/3682068}.

\bibitem[Maddison(2016)]{Maddison2016PoissonMC}
C.~J. Maddison.
\newblock A poisson process model for monte carlo.
\newblock \emph{arXiv preprint arXiv:1602.05986}, 2016.

\bibitem[Maddison et~al.(2014)Maddison, Tarlow, and Minka]{MaddisonTarlowMinka2014}
C.~J. Maddison, D.~Tarlow, and T.~Minka.
\newblock A* sampling.
\newblock In \emph{Advances in Neural Information Processing Systems 27 (NeurIPS 2014)}, pages 3086--3094. Curran Associates, Inc., 2014.

\bibitem[Mironov(2017)]{Mironov2017RDP}
I.~Mironov.
\newblock R\'enyi differential privacy.
\newblock In \emph{2017 IEEE 30th Computer Security Foundations Symposium (CSF)}, pages 263--275. IEEE, 2017.
\newblock \doi{10.1109/CSF.2017.11}.

\bibitem[Papandreou and Yuille(2011)]{PapandreouYuille2011}
G.~Papandreou and A.~L. Yuille.
\newblock Perturb-and-map random fields: Using discrete optimization to learn and sample from energy models.
\newblock In \emph{2011 International Conference on Computer Vision (ICCV)}, pages 193--200, Barcelona, Spain, Nov. 2011. IEEE.
\newblock ISBN 978-1-4577-1101-5.
\newblock \doi{10.1109/ICCV.2011.6126242}.
\newblock URL \url{https://home.ttic.edu/~gpapan/pubs/confr/PapandreouYuille_PerturbAndMap_ieee-c-iccv11.pdf}.

\bibitem[Patil et~al.(2024)Patil, Zhang, Wang, and Gonzalez]{Patil2024Gorilla}
S.~G. Patil, T.~Zhang, X.~Wang, and J.~E. Gonzalez.
\newblock Gorilla: Large language model connected with massive apis.
\newblock In \emph{Advances in Neural Information Processing Systems (NeurIPS)}, 2024.
\newblock URL \url{https://proceedings.neurips.cc/paper_files/paper/2024/file/e4c61f578ff07830f5c37378dd3ecb0d-Paper-Conference.pdf}.

\bibitem[Pearl(1984)]{Pearl1984}
J.~Pearl.
\newblock \emph{Heuristics: Intelligent Search Strategies for Computer Problem Solving}.
\newblock Addison-Wesley, Reading, MA, 1984.

\bibitem[Resnick(1992)]{Resnick1992}
S.~I. Resnick.
\newblock \emph{Adventures in Stochastic Processes}.
\newblock Birkh{\"a}user, Boston, 1992.

\bibitem[Reynolds(2020)]{Reynolds2020UniformFloat}
M.~B. Reynolds.
\newblock Uniform floating-point randoms.
\newblock \url{https://marc-b-reynolds.github.io/math/2020/06/16/UniformFloat.html}, 2020.

\bibitem[Schick et~al.(2023)Schick, Dwivedi-Yu, Dess{\`i}, Raileanu, Lomeli, Zettlemoyer, Cancedda, and Scialom]{Schick2023Toolformer}
T.~Schick, J.~Dwivedi-Yu, R.~Dess{\`i}, R.~Raileanu, M.~Lomeli, L.~Zettlemoyer, N.~Cancedda, and T.~Scialom.
\newblock Toolformer: Language models can teach themselves to use tools.
\newblock \emph{arXiv preprint arXiv:2302.04761}, 2023.

\bibitem[Xiao-Yang~Liu et~al.(2023)Xiao-Yang~Liu, Zha, Gao, Zhong, White, and Qiu]{dplora2312}
R.~Z. Xiao-Yang~Liu, D.~Zha, J.~Gao, S.~Zhong, M.~White, and M.~Qiu.
\newblock Differentially private low-rank adaptation of large language model using federated learning.
\newblock \emph{arXiv preprint arXiv:2312.17493}, 2023.

\bibitem[Yao et~al.(2023)Yao, Zhao, Yu, Du, Shafran, Narasimhan, and Cao]{Yao2023ReAct}
S.~Yao, J.~Zhao, D.~Yu, N.~Du, I.~Shafran, K.~Narasimhan, and Y.~Cao.
\newblock React: Synergizing reasoning and acting in language models.
\newblock In \emph{International Conference on Learning Representations (ICLR)}, 2023.
\newblock URL \url{https://openreview.net/forum?id=WE_vluYUL-X}.

\bibitem[Yu et~al.(2022)Yu, Naik, Backurs, Gopi, Inan, Kamath, Kulkarni, Lee, Manoel, Wutschitz, Yekhanin, and Zhang]{Yu2022DPFineTuneLM}
D.~Yu, S.~Naik, A.~Backurs, S.~Gopi, H.~A. Inan, G.~Kamath, J.~Kulkarni, Y.~T. Lee, A.~Manoel, L.~Wutschitz, S.~Yekhanin, and H.~Zhang.
\newblock Differentially private fine-tuning of language models.
\newblock In \emph{International Conference on Learning Representations (ICLR)}, 2022.
\newblock URL \url{https://openreview.net/pdf?id=Q42f0dfjECO}.

\bibitem[Zhu et~al.(2022)Zhu, Dong, and Wang]{ZhuDongWang2022FourierAccountant}
Y.~Zhu, J.~Dong, and Y.~Wang.
\newblock Optimal accounting of differential privacy via characteristic function.
\newblock In \emph{Proceedings of the 25th International Conference on Artificial Intelligence and Statistics (AISTATS)}, volume 151 of \emph{Proceedings of Machine Learning Research}, pages 4782--4817, 2022.
\newblock URL \url{https://proceedings.mlr.press/v151/zhu22c.html}.

\end{thebibliography}
